\documentclass[reprint, amsmath,amssymb, aps, prl ]{revtex4-1}

\usepackage{graphicx}
\usepackage{dcolumn}
\usepackage{bm}
\usepackage{color, xcolor, diagbox}
\usepackage[mathscr]{euscript}
\usepackage[colorlinks]{hyperref}
\hypersetup{
 colorlinks=true,
 citecolor=blue,
 linkcolor=teal,
 urlcolor=cyan}

\newcommand{\xin}{\hat{x}^\textrm{in}}
\newcommand{\qin}{\hat{q}^\textrm{in}}
\newcommand{\pin}{\hat{p}^\textrm{in}}
\newcommand{\ain}{\hat{a}^\textrm{in}}
\newcommand{\aout}{\hat{a}^\textrm{out}}
\newcommand{\xout}{\hat{x}^\textrm{out}}
\newcommand{\qout}{\hat{q}^\textrm{out}}
\newcommand{\pout}{\hat{p}^\textrm{out}}
\newcommand{\Lin}{\bm{L}^\textrm{in}}
\newcommand{\Lout}{\bm{L}^\textrm{out}}
\newcommand{\Ain}{\hat{A}^\textrm{in}}
\newcommand{\Aind}{\hat{A}^{\textrm{in}\dag}}
\newcommand{\ah}{\hat{a}}
\newcommand{\ad}{\hat{a}^\dag}
\newcommand{\USQ}{\hat{\mathcal{U}}_\textrm{sQ}}
\newcommand{\vac}{\textrm{vac}}
\newcommand{\Tr}[1]{\textrm{Tr}\left\{ #1\right\}}
\newcommand{\Tro}[1]{\textrm{Tr}_1\left\{ #1\right\}}
\newcommand{\Troe}[1]{\textrm{Tr}_{1,E}\left\{ #1\right\}}
\newcommand{\bra}[1]{\langle #1|}
\newcommand{\ket}[1]{|#1\rangle}

\newcommand{\add}[1]{#1}

\begin{document}

\title{\add{High-fidelity} bosonic quantum state transfer using imperfect transducers and interference}

\author{Hoi-Kwan Lau}
 \email{hklau.physics@gmail.com}
\author{Aashish A. Clerk}

\affiliation{Institute for Molecular Engineering, 
University of Chicago, 
5640 South Ellis Avenue, 
Chicago, Illinois 60637, U.S.A.}

\date{\today}

\begin{abstract}
We consider imperfect two-mode bosonic quantum transducers that cannot completely transfer an initial source-system quantum state due to insufficient coupling strength or other \add{Hamiltonian} non-idealities.    We show that such transducers can generically be made perfect by using interference and phase-sensitive amplification.  Our approach is based on the realization that a particular kind of imperfect transducer (one which implements a swapped quantum non-demolition (QND) gate) can be made into a perfect one-way transducer using feed-forward and/or injected squeezing.  We show that a generic imperfect transducer can be reduced to this case by repeating the imperfect transduction operation twice, interspersed with amplification.  Crucially, our scheme only requires the ability to implement squeezing operations and/or homodyne measurement on one of the two modes involved.  It is thus ideally suited to schemes where there is an asymmetry in the ability to control the two coupled systems (e.g.~microwave-to-optics quantum state transfer).  We also discuss a correction protocol that requires no injected squeezing and/or feed-forward operation.
\end{abstract}

\maketitle

\section{Introduction}

The ability to interface disparate quantum 
systems would allow one to harness their respective advantages, and could have a transformative effect on quantum science.  A crucial ingredient here is a quantum transducer: a device that faithfully transfers a quantum state from one system to another.
Quantum transduction is being actively pursued in a variety of settings \cite{2015PNAS..112.3866K}, including the conversion of quantum signals between microwave and optical frequencies \cite{2011JPhCS.264a2025R, 2012PhRvA..85b0302H, 2013NatPh...9..712B, 2014NatPh..10..321A, 2016PhRvB..93q4427H, Rueda:2016jg, 2016ApPhL.109c3107V, Higginbotham:2018ca}, the transfer of quantum information between a processor and memory \cite{Julsgaard:2004ud, Sherson:2006ju, Rabl:2006if, Wesenberg:2009es, 2010PhRvL.105v0501S, 2014PhRvX...4b1049G}, sympathetic cooling of mechanical oscillators \cite{2003PhRvL..90m7901M, 2010PhRvA..82b1803H, Lau:2014bv, 2015NatNa..10...55J}, and the generation of non-classical states for testing quantum mechanics and sensing \cite{1999JOptB...1..496P, 2003PhRvA..68a3808Z, 2015PhRvA..92e3804F, Reed:2017ci, 2017Sci...358..199C}.  

For a wide range of quantum systems, ideal transduction amounts to swapping the initial quantum states of two harmonic modes (e.g. photonic cavity modes or propagating temporal modes, spin ensembles, mechanical resonators).  Even in this setting, practical implementation is daunting.  Protocols are often limited by a weak interaction strength, a lack of full control, or unwanted spurious interactions.  In such cases, even without any injected environmental noise, one is only able to accomplish an ``incomplete" transduction where the desired input state is only partially transferred, and unwanted correlations between source and receiver are generated. Recent work has suggested creative strategies for mitigating errors in incomplete transducers. They however require the transducer to \add{be a quantum non-demolition (QND) gate \cite{Filip:2008gq, 2010PhRvA..81d2325M, Bennett:2016ks}} or have a very specific form \cite{2009PhRvA..80b2304F}, or require both full control and the ability to inject large amounts of squeezing at both \add{coupled systems} \cite{2018PhRvL.120b0502Z}.

In this paper, we describe and analyze an alternate approach to correcting imperfect quantum transducers that (unlike previous work) is fully general, and only requires control over {\it one} of the two coupled systems.  This makes it extremely attractive for, e.g., microwave to optics transduction, where control and squeezing is much easier for microwaves than optics (due to the toolbox of circuit QED \cite{Blais2004}).  The method is conceptually simple: apply the incomplete transduction operation twice, but intersperse them with a phase-sensitive amplification step.  Our main result is that interference in such a protocol can be used to cancel reflections via destructive interference, allowing an ideal (i.e.~complete) unidirectional state transfer.    
In the absence of environmental noise, this strategy allows for perfect transduction using imperfect transducers; even with such noise, it can still provide a marked advantage.
We also discuss a multi-pass interference protocol that can make an imperfect transducer perfect without any need to inject squeezing or perform measurement and feed-forward.

\begin{figure}
\begin{center}
\includegraphics[width=\linewidth]{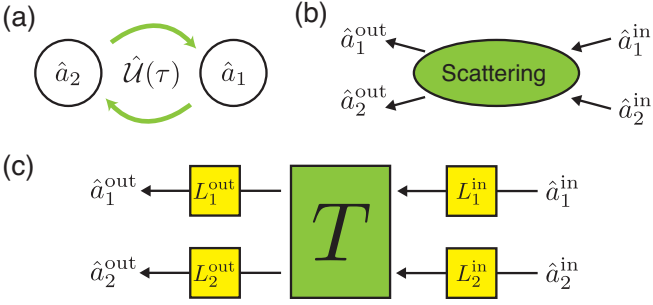}
\caption{{\bf Transduction between two bosonic modes.}  
(a) Intra-cavity transfer: two stationary modes evolve in time under the action of a coupling Hamiltonian.    
(b) Scattering scenario: input and output propagating modes interact through scattering.
(c) General picture: transduction is a two-mode linear unitary transformation (green).  To improve operation, one can apply single-mode transformations (yellow) to each mode both before and after the transduction operation. 
\label{fig:intro}}
\end{center}
\end{figure}

\section{Results}

\textbf{Classification.}  We start by considering the simplest kind of transducer, where the interaction of two bosonic modes $\hat{a}_1$, $\hat{a}_2$ is described by a Gaussian unitary transformation; the additional effect of environmental noise will be treated later.  Depending on the setup, our transducer could be implemented via time evolution under a Hamiltonian which couples the two systems (Fig.~\ref{fig:intro}a), or via a scattering process involving propagating modes (Fig.~\ref{fig:intro}b).  
For concreteness, we will use throughout the nomenclature of a scattering process;  the results are
easily applied to the time-domain case by realizing that input (output) modes correspond to initial (final) time mode operators, i.e.~$\ain \equiv \hat{a}(0)$ and $\aout \equiv \hat{a}(\tau) = \hat{\mathcal{U}}^\dag(\tau) \hat{a}(0) \hat{\mathcal{U}}(\tau)$ for some evolution operator $\hat{\mathcal{U}}$ corresponding to an evolution time $\tau$.

In this setting, an ideal transducer simply swaps the state of the two modes, i.e. $\hat{a}^\textrm{out}_{2} = \hat{a}_1^\textrm{in}$, $\hat{a}^\textrm{out}_1 = \hat{a}_2^\textrm{in}$.  Any device with a more complicated input-output relation is an ``incomplete" transducer.  
Introducing Hermitian quadrature operators via $\hat{a} \equiv (\hat{q}+i\hat{p})/\sqrt{2}$, our generic incomplete transducer corresponds to a transformation
\begin{equation}\label{eq:incomplete}
\left( \begin{array}{c} \qout_i \\ \pout_i \end{array}\right) = 
\sum_{j=1}^2\left( \begin{array}{cc} T^{qq}_{ij} & T^{qp}_{ij} \\ T^{pq}_{ij} & T^{pp}_{ij}\end{array}\right)\left( \begin{array}{c} \qin_j \\ \pin_j \end{array}\right)
\equiv \sum_{j=1}^2 \bm{T}_{ij} \left( \begin{array}{c} \qin_j \\ \pin_j \end{array}\right)~.
\end{equation}
Here $\bm{T}$ is a $4\times4$ real, symplectic matrix.  Its entries $T^{xy}_{ij}$ are scattering matrix elements in the quadrature basis 
(where $i, j \in \{1,2 \} $ and $x, y \in \{q,p \} $).  In particular, the $2\times 2$ sub-matrix $\bm{T}_{ij}$ describes the transmission/reflection of quadratures from mode $j$ to $i$.  
An incomplete transducer will in general have non-unity transmission and non-zero reflections.  This implies that the input state from the source system will be imperfectly transmitted to the target system, and also corrupted by fluctuations of the partially reflected incident state at the target.

Our first goal will be to usefully classify such incomplete unitary transducers.  The irreducible resource of interest is the ``non-locality" of the transducer, i.e.~its ability to make the two disparate modes interact \footnote{Note that the locality here is referred with respect to the bosonic degree of freedom, but the physical systems that contain the degree of freedom are not necessarily spatially separated.}.  We are thus interested in understanding what kinds of transformations are equivalent up to purely  local 
(i.e.~single-mode) transformations of the two modes (possibly both before and after the transduction operation, see Fig.~\ref{fig:intro}c).  Such local operations cause the reflection/transmission matrices to transform as $\bm{T}_{ij} \rightarrow   \bm{L}_i^\textrm{out} \bm{T}_{ij}\bm{L}_j^\textrm{in} \equiv \bm{T'}_{ij} $ where $\bm{L}^{\rm in (out)}_i$ is the $2\times2$ symplectic matrix describing the transformation of mode $i$ before (after) transduction \cite{Weedbrook:2012fe}.  One finds that such transformations always preserve matrix rank and determinant:  $\textrm{rank}(\bm{T}_{ij}) = \textrm{rank}(\bm{T'}_{ij})$, $\textrm{det}(\bm{T}_{ij}) = \textrm{det}(\bm{T'}_{ij})$.  We will thus use these quantities as the basis of our non-locality based classification.

The main element of the classification is matrix rank.  To see the physical meaning of these quantities, consider the singular-valued decompositions (SVD) of the reflection/transmission matrices:
\begin{equation}\label{eq:SVD}
\bm{T}_{ij} = \bm{V}_{ij}\bm{D}_{ij}\bm{W}_{ij}\equiv \bm{V}_{ij} \left(\begin{array}{cc} D^{(q)}_{ij} &0 \\ 0 & D^{(p)}_{ij} \end{array}\right) \bm{W}_{ij}~,
\end{equation}
where $\bm{V}_{ij}$ and $\bm{W}_{ij}$ are rotation matrices and $\bm{D}_{ij}$ is diagonal.  Physically, the rotation matrices effectively redefine the input and output quadratures \cite{Weedbrook:2012fe}, e.g. one can define a new $q$-quadrature as $\hat{q}'\equiv \cos \phi \hat{q} + \sin\phi \hat{p}$.  It is then clear that rank$(\bm{T}_{ij})$, which is the number of non-vanishing singular values in $\bm{D}_{ij}$, characterizes the number of effective quadratures that are transmitted/reflected from mode $j$ to $i$.  These quantities are invariant under local transformations.

While there are four transmission/reflection matrices $\bm{T}_{ij}$, the ranks of these matrices are not independent:  they are constrained by the requirement that scattering preserves canonical commutation relations  (see Methods for more detail):
\begin{eqnarray} 
&\textrm{rank}(\bm{T}_{ij}) = \textrm{rank}(\bm{T}_{\bar{i}\bar{j}})~,\label{eq:M_eq} \\
&\det(\bm{T}_{i1}) + \det(\bm{T}_{i2}) = 1~, \label{eq:M_det}
\end{eqnarray}
where $\bar{i}=1$ if $i=2$, and $\bar{i}=2$ if $i=1$.  Eq.~(\ref{eq:M_eq}) shows that the number of transmitted and reflected quadratures in both modes are the same.  We denote these integers by $n_\textrm{T}\equiv \textrm{rank}(\bm{T}_{21})$ and $n_\textrm{R} \equiv \textrm{rank}(\bm{T}_{22})$. Further, Eq.~(\ref{eq:M_det}) implies that at least one of $n_\textrm{T}$, $n_\textrm{R}$ must be equal to 2. The upshot is that we have five distinct classes of two-mode transformations (five possible choices of $[[n_\textrm{T},n_\textrm{R}]]$, see Table \ref{table:FullClass}).
The physical interpretation is as discussed above:  a class [[$n_\textrm{T},n_\textrm{R}$]] transducer will transmit $n_\textrm{T}$ quadratures and reflect $n_\textrm{R}$ quadratures.

We stress that two transducers from different classes cannot be made equivalent by applying local transformations only.  In contrast, if two transducers belong to the same class, and that class is not [[2,2]], then they are equivalent up to purely local transformations.  The situation is slightly more complicated for class [[2,2]]:  two transducers in this class can only be made equivalent via local transformations if they have the same value of $\chi \equiv \det(\bm{T}_{21})$, as the determinant cannot be altered by local operations alone.  This distinction for class [[2,2]] (and the determinant $\chi$) will not play a role in the transduction-correction strategy that we develop below.
 
 \begin{table}
\begin{center}
\begin{tabular}{|c|cc|c|c|}
\hline Class & $n_T$ & $n_R$ & $\chi\equiv \det(\bm{T}_{21})$ & Equivalent operation \\ \hline \hline
[[0,2]] &  0 & 2 & 0 & Identity \\ \hline
[[1,2]] &  1 & 2 & 0 & QND gate \\ \hline
~ & ~ & ~  & $0>\chi$ & Two-mode squeezing \\ \cline{4-5}
[[2,2]] & 2 & 2 & ~$1>\chi>0$~ & Beam splitter \\ \cline{4-5}
~ & ~ & ~ & $\chi>1$ & ~Swapped two-mode squeezing ~\\ \hline
[[2,1]] &  2 & 1 & 1 & Swapped QND gate \\ \hline
[[2,0]] &  2 & 0 & 1 & {\small SWAP} \\ \hline
\end{tabular}
\end{center}
\caption{Classification of two-mode linear transformations in terms of the rank ($n_{\rm T}$,$n_{\rm R}$) of transmission matrices and reflection matrices.  
Except class [[2,2]], all transformations in a given class are equivalent up to local transformations of the two modes.   Class [[2,2]] transformations are subject to an extra constraint: local transformations cannot change the determinant $\chi$ of the transmission matrix.  Note that the value of $\chi$ plays no role in determining a transformation's utility in our transduction-correction scheme.  
The last column shows the equivalent 
well-known operation for each class, as can be obtained via the quadrature-diagonalization procedure discussed in the main text (full details in Methods).
\label{table:FullClass}}
\end{table}


\textbf{Quadrature diagonal form.}  To develop a systematic strategy for correcting incomplete transducers, we find that it is useful to use local transformations (as depicted in Fig.~\ref{fig:intro}c) to make a given initial transducer \textit{quadrature-diagonal}:  the scattering of the composite system does not mix $\hat{q}$ and $\hat{p}$ quadratures.  This implies that all four reflection/transmission matrices become diagonal, i.e.
\begin{equation}
	\bm{T}_{ij} \rightarrow \Lout_i \bm{T}_{ij}\Lin_j = \textrm{diag}\left(\Lambda^{(q)}_{ij}, \Lambda^{(p)}_{ij} \right)~.
	\label{eq:diagonalize}
\end{equation}
As we show explicitly in the Methods, one can always construct the needed local transformations
$\bm{L}^{\rm in/out}_i$.  This diagonalization procedure also helps provide intuition, as it shows that each class of transducer is equivalent to a well-known two-mode Gaussian operation (see Table~\ref{table:FullClass}), e.g.~beam-splitters (BS), two-mode squeezers (TMS), and QND gates.

While local operations always exist to make our transducer quadrature-diagonal, in general these operations will require the ability to implement single-mode squeezing operations for {\it both} systems $1$ and $2$.  In many realistic situations (e.g.~microwave to optics transfer), squeezing can only be implemented on one mode (say mode $1$).  With this additional constraint, it is no longer possible in general to locally transform the scattering into a fully quadrature diagonal form.  Nevertheless, as shown in the Methods, one can still have three of the four output quadratures be in diagonal form (i.e.~they exhibit no mixing of $q$'s and $p$'s).  This more limited kind of quadrature diagonal form will be sufficient for our correction scheme.


\textbf{Good transducer classes.}  For transduction, the ideal case is clearly [[$2,0$]]:  transmission of both quadratures, no reflections at all.  After quadrature-diagonalization, this class of transducer can always be converted to a perfect {\small SWAP}.  The classes other than [[$2,0$]] will be imperfect transducers.

We first discuss imperfect transducers that belong to class [[$2,1$]].  While such transducers effectively transmit two quadratures, they also have non-trivial reflections; such reflections would cause unwanted correlations between source and receiver and would ultimately degrade the transfer fidelity.  This class corresponds to a partial impedance matching: only one quadrature is impedance matched and transmitted without reflections.

We show here that such an imperfect transducer can be easily corrected {\it if} one is content with unidirectional transduction, meaning that at the end of the operation, $\hat{a}^{\rm out}_2 = \hat{a}^{\rm in}_1$, but  $\hat{a}^{\rm out}_1 \neq \hat{a}^{\rm in}_2$.  Such one-way transduction is more than sufficient in many applications, where one ultimately wants to transfer a state from source to target, and does not care about the final state of the source.  As we elaborate below, the key observation is that for a system in class [[2,1]], the imperfect impedance matching in one quadrature can be remedied either by injecting squeezing, or via a measurement-plus-feedback operation.

To see this, we first put our class [[2,1]] incomplete transducer into quadrature-diagonal form by implementing appropriate local operations (c.f.~Eq.~(\ref{eq:diagonalize})).  We stress this diagonalization can be accomplished without needing the ability to locally squeeze both modes $1$ and $2$.  With the appropriate local transformations, the scattering relations take the form:
\begin{eqnarray}\label{eq:SQND_quad}
	\qout_1 = -\eta \qin_1 + \qin_2  ~&,&~\qout_2 = \qin_1~, \nonumber \\
	\pout_1 = \pin_2 ~&,&~\pout_2 = \pin_1 + \eta \pin_2~,
\end{eqnarray} 
where $\eta$ is the singular value of the original reflection matrix $\bm{T}_{22}$ (see Methods for details).  As expected for class [[2,1]], all input quadratures are transmitted to the opposite mode, while for each system, there is one quadrature that is partially reflected with amplitude $\pm \eta$.

This transformation corresponds to the action of the unitary operator
\begin{equation}\label{eq:SQND_gate}
	\USQ(\eta) \equiv  \hat{\mathbb{S}} \exp(i\eta \hat{q}_1 \hat{p}_2)= \exp(i\eta \hat{p}_1 \hat{q}_2) \hat{\mathbb{S}} ~,
\end{equation}
which is the product of a {\small SWAP} operator $\hat{\mathbb{S}} \equiv \exp\left(i \frac{\pi}{2} (\ad_1 - \ad_2)(\ah_1 - \ah_2)\right) $, and a QND gate.  While the SWAP operator generates the ideal transduction operation, the QND operation creates unwanted correlations between the two systems in one quadrature. We refer this less-known composite operation as a ``swapped QND" gate, or simply sQND.

For one-way transduction, we only care about the initial state of the source system, and the final state of the target system.  One can then use remaining input and output mode to undo the imperfections created by the unwanted QND gate \cite{2006PhRvL..97k0501M, Gu:2009ed}.  For example, if the desired transduction is from mode 2 to 1, the unwanted reflection of the $\qin_1$ quadrature can be mitigated by injecting a squeezed state into mode 1, such that $\qin_1\rightarrow 0$ in Eq.~(\ref{eq:SQND_quad}).  In contrast, if the desired transduction is from mode $1$ to $2$, the unwanted reflection of the $\pin_2$ quadrature can be corrected by making a homodyne measurement of $\pout_1$, and using the fact $\pout_1 = \pin_2 $.  The unwanted contamination of $\pout_2$ by $\pin_2$ in Eq.~(\ref{eq:SQND_quad}) will become a phase-space displacement proportional to the measurement outcome $p_1$; it can thus be corrected by applying an appropriate compensating displacement to mode 2.
\add{We note that the strategy of mitigating error by homodyne detection and injected squeezing is reminiscent of the protocol in Ref.~\cite{2018PhRvL.120b0502Z}.  There is an important difference however:  as a sQND operation reflects only one unwanted quadrature, either squeezing of homodyne is sufficient for correction.  In contrast, both these operations are needed in Ref.~\cite{2018PhRvL.120b0502Z}.}


\textbf{Bad transducer classes.}  
We now turn to the remaining imperfect classes of transducers.  Class [[0,2]] has zero transmission between the two systems, and essentially describes two uncoupled systems; there is thus no way to fix it.  In contrast, the remaining imperfect classes [[2,2]] and [[1,2]] have non-zero transmission and coupling between the two systems.  They however have no impedance matching: reflections involve both quadratures, and the trick used to ameliorate class [[2,1]] cannot be directly applied.

\begin{figure}
\begin{center}
\includegraphics[width=\linewidth]{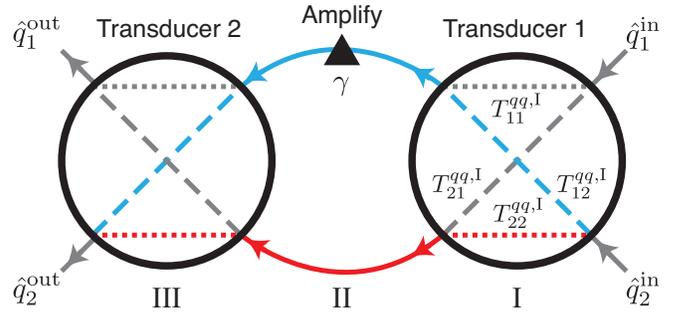}
\caption{{\bf Schematic depicting the intereference-approach to transduction.}  
Two imperfect transducers (from class [[2,2]] or [[1,2]], denoted by circles) are operated sequentially, interspersed by a phase-sensitive amplification step (triangle).  As discussed in the text, local transformations can
be made to ensure $q$ quadrature scattering is independent of $p$ quadratures.  By adjusting the amplification strength $\gamma$, the two pathways for reflection (blue and red) can be made to interfere destructively.  As a result, we have a partial impedance matching:  $\qout_2$ only depends on $\qin_1$.
\label{fig:interfere}}
\end{center}
\end{figure}

Despite the lack of any impedance matching, there is indeed a way to make transducers in these remaining imperfect classes
perfect for one-way transduction.  Our approach is to use interference to cancel reflections in one quadrature, making the system equivalent to class [[2,1]] (where one quadrature is impedance matched).  This can be accomplished by simply using the imperfect transducer twice (or more generally, using two distinct imperfect transducers sequentially).  
Our protocol involves three stages: (I) feeding the input modes into the first bad transducer; (II) applying a carefully tuned single-mode phase-sensitive amplification (squeezing) on mode 1; (III) feeding both modes into the second bad transducer (see Fig.~\ref{fig:interfere}).  We note that the two bad transducers need not be identical, nor even in the same class (though they could be). 

We start by using local transformations (c.f.~Eq.~(\ref{eq:diagonalize})) to make both imperfect transducers $q$-quadrature diagonal (i.e.~the $\qout_j$ are only functions of the $\qin_{j'}$).   With these local transformations, the output $q$-quadratures of each transducer are described by:
\begin{equation}\label{eq:T_matrix}
\left( \begin{array}{c} \qout_1 \\ \qout_2 \end{array}\right) =   
\left( \begin{array}{cc}T^{qq,\alpha}_{11} & T^{qq,\alpha}_{12} \\ T^{qq,\alpha}_{21} & T^{qq,\alpha}_{22}  \end{array}\right)\left( \begin{array}{c} \qin_1 \\ \qin_2 \end{array}\right)\equiv
\bm{T}^{qq,\alpha}\left( \begin{array}{c} \qin_1 \\ \qin_2 \end{array}\right)
\end{equation}
where $\alpha \in$\{I,III\} indexes the two imperfect transducers, and $\bm{T}^{qq,\alpha}$ is a sublock of each transducer's scattering matrix (c.f.~Eq.~(\ref{eq:incomplete})).    The fact that the $q$ quadrature scattering is independent of $p$ quadratures means that we can consider them alone in what follows.  

We will next concatenate the two transducers, with a phase-sensitive amplification step in between, such that for the composite system, $\qout_2 \propto \qin_1$.  The amplification step corresponds carefully tuned squeezing or anti-squeezing along the $q$ direction in phase space, i.e. $\qout = \gamma \qin$, where $\gamma$ is an arbitrary real number describing the squeezing / anti-squeezing strength (and a possible phase shift).  The composite scattering process (transduction, amplification, transduction) is illustrated in Fig.~\ref{fig:interfere}.  
The overall transformation of $\qout_2$ is
\begin{eqnarray}
\qout_2 &=& (T^{qq,\textrm{III}}_{22}T^{qq,\textrm{I}}_{21} + \gamma T^{qq,\textrm{III}}_{21}T^{qq,\textrm{I}}_{11}) \qin_1 \nonumber \\
&& + (T^{qq,\textrm{III}}_{22}T^{qq,\textrm{I}}_{22} + \gamma T^{qq,\textrm{III}}_{21}T^{qq,\textrm{I}}_{12}) \qin_2~.
\label{eq:interference}
\end{eqnarray}
The last line of this equation shows that the reflection of $\qin_2$ involves the coherent sum of two processes: reflection from each transducer (first term), or two transmission events interspersed with amplification (second term).

We can now achieve our goal of impedance matching the $q$ quadrature at output mode $2$:  reflections can be cancelled by picking $\gamma$ to cause destructive interference between the two possible pathways.  One needs: 
\begin{equation}\label{eq:gamma}
	\gamma = -T^{qq,\textrm{III}}_{22}T^{qq,\textrm{I}}_{22}/T^{qq,\textrm{III}}_{21}T^{qq,\textrm{I}}_{12}~.  
\end{equation}
With this choice of $\gamma$, the resulting ``composite" transducer has no $q$ quadrature reflections at output mode $2$.  As a result, the composite system's reflection matrix $\bm{T}_{22}$ has at least one zero eigenvalue, and hence is at most rank 1.  The composite system is thus necessarily a transducer in class [[2,1]] or [[2,0]] (c.f. Eq.~(\ref{eq:M_det})).  As already discussed, by simply using additional purely local operations, we can now use this composite system for perfect one-way transduction.  We stress that these local operations need not require any mode-$2$ squeezing.  We also stress that because two quadratures are transmitted in these two bad classes, it is always possible to redefine quadratures such that both transmission amplitudes contributing to $\hat{q}_2$ reflection (i.e. $T^{qq,\textrm{III}}_{21}$ and $T^{qq,\textrm{I}}_{12}$) are non-vanishing.

\textbf{Rapid state transfer.}  As an example of a practical application, we show how our protocol can be used to accelerate state transfer between a quantum processor (mode $1$) and a quantum memory (mode $2$).  In many physical platforms (see e.g.~\cite{Rabl:2006if, Wesenberg:2009es}), the state transfer is achieved by evolving under a tunneling interaction $\hat{H}_\textrm{BS}\equiv ig(\hat{a}^\dag_1 \hat{a}_2 - \hat{a}_1 \hat{a}^\dag_2)$ for a time $t$.  The input-output transformation has the form of a BS transformation, i.e.
\begin{equation}\label{eq:1BS}
	\xout_1 = \cos\theta \xin_1 + \sin\theta \xin_2~~,~~\xout_2 = -\sin\theta \xin_1 + \cos\theta \xin_2~,
\end{equation}
where $\hat{x}$ is either quadrature, and the BS angle is $\theta \equiv gt$.   One only has a complete state transfer when the evolution time reaches $t=\tau_0 \equiv \pi/2g$; for this choice, the BS transformation is a perfect {\small SWAP}.  In many relevant cases, the coupling $g$ is weak, making the transfer process exceedingly slow. 
If one instead uses a time $t < \tau_0$,  we have an incomplete transducer belonging to class $[[2,2]]$.

By now using our interference-based scheme (see Fig.~\ref{fig:interfere}), the state transfer time $\tau$ can be dramatically shortened, i.e. $\tau \ll \tau_0$.  Explicitly, the tunneling interaction is first applied for a time $\tau/2$, then the processor is squeezed, and then finally the tunneling interaction is applied for another period $\tau/2$.  For each round of tunneling, a BS with $\theta = g\tau/2$ is implemented; the effective transmission amplitude is $T^{qq}_{12}=-T^{qq}_{21}=\sin\theta$ and reflection amplitude is $T^{qq}_{22}=\cos\theta$.  According to Eq.~(\ref{eq:gamma}), $\qout_2$ is impedance matched if $\gamma = \cot^2\theta$.  With this choice, the overall transformation is given by
\begin{eqnarray}
\qout_1 = \eta \cot^2 \theta \qin_1 + \cot \theta \qin_2 ~~,~~ \qout_2 = -\cot \theta \qin_1 ~,\nonumber \\
\pout_1 = \tan \theta \pin_2 ~~,~~ \pout_2 = -\tan \theta \pin_1 +\eta  \pin_2~,~
\label{eq:TwoBS}
\end{eqnarray}
where $\eta = 1-\tan^2\theta$.  This transformation belongs to class [[2,1]] because each output mode consists of two transmitted and one reflected quadratures.  

We can now use the previously discussed strategies to make our composite class [[2,1]] transducer perfect.  For a memory write-in (transfer from processor mode-$1$ to memory mode-$2$), the reflection noise can be eliminated by measuring the processor and post-processing.  The state is then faithfully transferred if the input state is first squeezed by $- \tan \theta$ along the $q$-quadrature before being sent into the composite transducer.  Similarly, for a memory readout, the reflection noise is eliminated by initially preparing the processor in a squeezed state; the state transfer is then faithful if the output state leaving the composite transducer is squeezed  by $\tan\theta$ along the $q$-quadrature.  
In principle, our protocol can {\it arbitrarily} reduce the hopping time,
\add{provided that a sufficiently strong degree of squeezing $\gamma$ is available (i.e. $\tau\sim \sqrt{1/\gamma} \tau_0$). }

\add{Although the above discussion focuses on time-domain transduction, the same principle applies to scattering-mode transducers whose scattering matrix corresponds to an incomplete BS operation (i.e. $\theta < \pi/2$ in Eq.~(\ref{eq:1BS})).  For example, an impedance-mismatched microwave-optical transducer is effectively a partial BS in the resolved sideband regime.  Our scheme can be applied by collecting both microwave and optical output after first round of transduction, then parametrically amplifying the microwave field, and finally feeding both fields back to the transducer again.  If the optical input is prepared in a squeezed state, or the microwave output is homodyne detected, the transduction can be completed even though the transducer is impedance-mismatched.
}

We also note that apart from accelerating the state transfer that is mediated by tunneling interaction, in Methods we show a counterintuitive example that our scheme can also implement a perfect state transfer even if the modes are coupled only through two-mode-squeezing interaction.

\textbf{Lossy transduction.}  While the advantage of our approach for such accelerated memory transfer is clear in the case of purely unitary evolution, it also provides an advantage in the more realistic case where there is loss and injected noise from an external environment.
To illustrate this idea, we study a generic scenario where the processor mode is lossy (subject to damping and injected noise), whereas the memory mode is essentially lossless.  The system evolution follows the Heisenberg-Langevin equation
\begin{equation}\label{eq:LBS}
\dot{\hat{a}}_1 = -(\kappa/2) \hat{a}_1 + g \hat{a}_2 -i \sqrt{\kappa} \hat{B}^\textrm{in} ~~,~~\dot{\hat{a}}_2 = -g \hat{a}_1~,
\end{equation}
where $\kappa$ is the mode-$1$ loss rate, and $\hat{B}^\textrm{in}$ denotes incident vacuum noise from the bath. 

\begin{figure}
\begin{center}
\includegraphics[width=\linewidth]{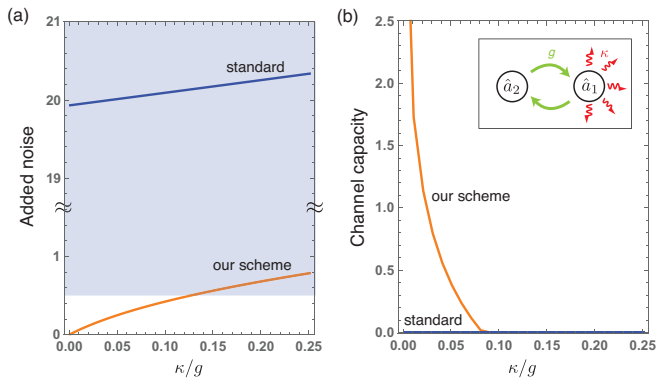}
\caption{{\bf Accelerated beam-splitter state transfer.}  A quantum state is transferred from a lossy processor mode $\hat{a}_1$ into a quantum memory mode $\hat{a}_2$ by using a tunneling interaction (strength $g$), c.f.~Eq.~(\ref{eq:LBS}) (illustration in inset).  We restrict the protocol time to be ten times shorter than the time required for a perfect swap, $\tau=\tau_0/10 = \pi / 20 g$.  
(a) Added noise $\overline{\mathcal{N}}_{\min}$ in the transduced state, as a function of processor loss $\kappa$.  The added noise of the interference scheme (orange) is dramatically smaller than that of the standard approach (blue).  The shaded area, $\overline{\mathcal{N}}_{\min}\geq 0.5$, shows the added noise that would result if transduction was attempted using a classical measurement plus feedforward strategy \cite{Braunstein:2005wr}; anything below this level thus represents quantum transduction.  Our scheme is below this boundary for $\kappa \lesssim 0.1 g$.  
(b) Quantum channel capacity for the interference approach (orange) and the standard approach (blue).  The result shown is a lower bound for our scheme \cite{2001PhRvA..63c2312H, 2009PhRvL.102e0503P} (see Methods for details) but exact for the standard approach.  The capacity of the standard approach is rigorously zero even without loss.  In contrast, our interference scheme yields finite capacity when $\kappa \lesssim 0.1 g$.  Both results suggest that the our scheme enables rapid quantum transduction even in the presence of loss.
\label{fig:noise6}}
\end{center}
\end{figure}

Without loss of generality, we consider a fast ($\tau \ll \tau_0$) write-in process.  Our 3-step interference scheme is compared against the standard approach, where the system evolves as per Eq.~(\ref{eq:LBS}) for a time $\tau \ll \tau_0$, and the processor mode is discarded at the end of the protocol.
We quantify the performance of these schemes by considering the total added noise of the protocol, expressed as an equivalent number of noise quanta $\mathcal{N}$ added to the initial mode-$1$ state.  This metric has been used in recent experiments \cite{Higginbotham:2018ca}, and is analogous to how one quantifies the performance of quantum amplifiers \cite{1982PhRvD..26.1817C}.  
As a typical example, we consider a write-in process that is ten-times faster than a conventional state transfer (i.e. $\tau=\tau_0/10$).  As shown in Fig.~\ref{fig:noise6}a, the added noise of our scheme is an order of magnitude less than the standard approach even when loss is significant, $\kappa \approx 0.1 g$.

While added noise is a convenient experimental metric, it is also interesting to consider the performance of our transducer as a quantum channel: to what extent can it be used to transfer quantum information?  By treating one-way transduction as an effective single-mode Gaussian channel \cite{2018PhRvL.120b0502Z}, the transducer's performance can be quantified by the quantum channel capacity, which describes the number of qubits that can be reliably sent per use of the transducer (asymptotically in the case where the transducer is used many times) \cite{Weedbrook:2012fe}. 
As shown in Fig.~\ref{fig:noise6}b, for $\tau = \tau_0 / 10$, the channel capacity of the standard approach vanishes even in the absence of loss.  It is because the transmission amplitude is smaller than the reflection, so the channel is antidegradable and thus incapable of tranducing quantum information \cite{2006PhRvA..74f2307C}.  
In contrast, our approach is able to yield a non-zero channel capacity using the same interaction strength $g$ and total evolution time (see Methods for details).  This is true even when there are moderate levels of loss,   $\kappa \lesssim 0.1 g$.

\textbf{Imperfect detection or input squeezing.}  The discussion of our correction strategy has so far relied on making an imperfect transducer setup equivalent to a
class [[2,1]] sQND operation, and then eliminating the unwanted QND interaction either by making a perfect homodyne measurement, or by injecting infinite squeezing.  In any realistic setting, neither of these last two operations will be implemented perfectly.  
The resulting errors would however not be completely detrimental to the transduction: they would lead to the incomplete suppression of added noise in only one of two canonical quadratures (e.g. the $p$ quadrature noise of output mode $2$ in Eq.~(\ref{eq:SQND_quad})).   The other conjugate quadrature would  still be transduced perfectly.  

Imperfect transduction where there is only added noise in one quadrature of the transduced state (i.e. class [[2,1]]) can be exploited in many important applications.  For example, if we are only interested in transferring a bosonic-encoded qubit \cite{2018PhRvA..97c2346A}, 
protocols for eliminating added noise in one quadrature have already been proposed  
(c.f.~supplementary Sec.~VII in Ref.~\cite{Higginbotham:2018ca} \add{and a more general strategy in \cite{2018PhRvA..97c2346A}}.)  
Furthermore, we show in Methods that by using squeezed versions of existing encodings of logical qubits \cite{2013PhRvA..87d2308F, LeJeannic:2018ek}, or by using noiseless subsystems \cite{Knill:2000wt, Zanardi:2001vo, Kempe:2001bg, 2017PhRvA..95b2303L, Marshall2018}, the effects of single-quadrature transduction noise can also be mitigated \add{without conducting active error correction}.

An alternate and perhaps more elegant approach to completely eliminate the requirement of homodyne measurement or injected squeezing is to simply apply more than two rounds of incomplete transduction.  
Our approach requires three sQND operations interspersed with amplification.  Each sQND can be implemented by quadrature-diagonalizing a class [[2,1]] transducer, or constructed by two class [[1,2]] or [[2,2]] transducers as described above (c.f.~Eq.~(\ref{eq:interference})).  The net result is that one needs at most six incomplete transduction operations.  \add{We note that such construction of a perfect transducer can be further simplified if each incomplete transduction is restricted to a specific form \cite{2015PhRvA..92b2346K, Bennett:2016ks}.}

By controlling each amplification strength interspersing the sQND operations, the strength of the unwanted QND gates will be effectively adjusted \cite{H_amp}.  Our strategy is to adjust each QND gate strength, such that the net effect of the three QND operations cancel.  One is just left with the {\small SWAP} part of each of the three sQND operations; as three {\small SWAP}'s is a {\small SWAP}, the  overall transformation becomes a pure {\small SWAP}, and thus described a perfect (two-way) transduction. 

\begin{figure}
\begin{center}
\includegraphics[width=\linewidth]{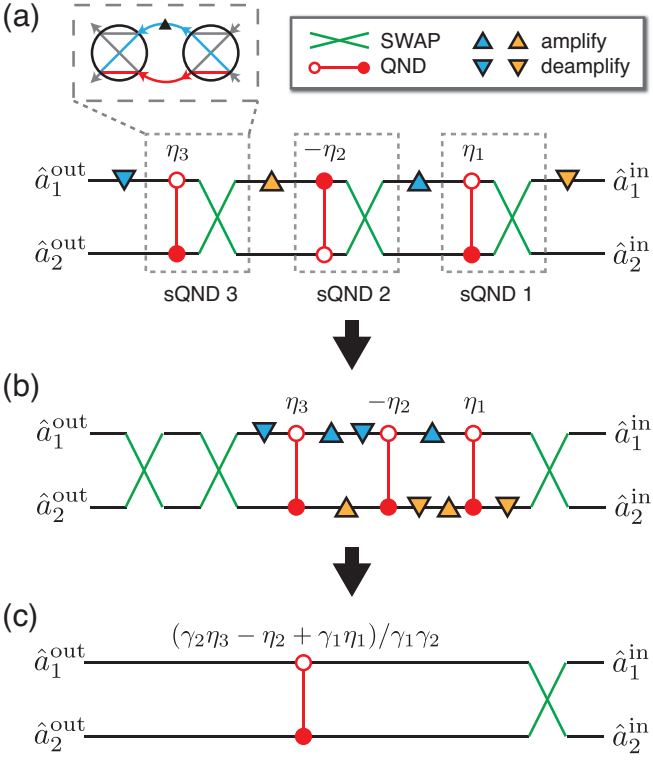}
\caption{ \label{fig:6pass} {\bf Circuit diagram for performing perfect two-way transduction.} 
Using (at most) six imperfect transduction steps, no input squeezing or measurement is required. (a) Circuit diagram of Eq.~(\ref{eq:6pass}), which involves three sQND (dotted box).  Each sQND can be constructed by one class [[2,1]] transducer, or two bad transducers (dashed box).  
Red dumbbell represents QND gate, $\exp(i\eta \hat{x}_1 \hat{y}_2)$, where $x,y=q$ ($p$) for solid (hollow) circle.  The QND strength is shown above.  
Blue (orange) triangles are amplification/deamplification with strength $\gamma_1$ ($\gamma_2$).  (b) Equivalent circuit after gate rearrangement.  Pairs of amplification-deamplification are added by using the fact that $\hat{G}^\dag(\gamma)\hat{G}(\gamma)=\hat{G}(\gamma)\hat{G}^\dag(\gamma)=\hat{I}$.  The amplification gates effectively scale the QND gate strength as $\eta_1 \rightarrow (1/\gamma_2)\eta_1$, $\eta_2 \rightarrow (1/\gamma_1\gamma_2)\eta_2$, $\eta_3 \rightarrow (1/\gamma_1)\eta_3$  (details in Methods).  (c)  After removing redundant {\small SWAP} and merging the QND gates, the resultant circuit becomes a single swapped QND gate with adjusted QND strength.}
\end{center}
\end{figure}

More precisely, the sequence of local transformations and incomplete transduction steps required are described by the following composite operator (see circuit model in Fig.~\ref{fig:6pass}):
\begin{eqnarray}\label{eq:6pass}
&&\hat{G}^\dag(\gamma_1) \USQ(\eta_3) \hat{G}(\gamma_2)\USQ^\dag(\eta_2) \hat{G}(\gamma_1) \USQ(\eta_1) \hat{G}^\dag(\gamma_2) \nonumber \\
&=& \USQ\left(\frac{\gamma_1 \eta_1 -\eta_2 +\gamma_2 \eta_3}{\gamma_1 \gamma_2}\right)~.
\label{eq:SixTimes}
\end{eqnarray}
Here, the unitary operator $\USQ(\eta)$ describes a sQND gate with QND strength $\eta$, and is defined in Eq.~(\ref{eq:SQND_gate}).
We also use $\hat{G}(\gamma)$ to denote a single-mode squeezing operation on mode-$1$ that transforms $\hat{q}_1 \rightarrow \gamma \hat{q}_1$, $\hat{p}_1 \rightarrow (1/\gamma) \hat{p}_1$.  
Note that the middle pair of transduction operations needs to correspond to $\USQ^\dag$.  This can be implemented by simply interchanging the definition quadratures, i.e. $\hat{q} \rightarrow \hat{p}$ and $\hat{p} \rightarrow -\hat{q}$, before and after a sQND.  

As seen from Fig.~\ref{fig:6pass}, the composite operation described by Eq.~(\ref{eq:SixTimes}) is equivalent to a single sQND with a modified QND gate strength.
Finally, if we pick $\gamma_1$ and $\gamma_2$ to satisfy $\eta_2 = \gamma_1 \eta_1  +\gamma_2 \eta_3$, 
the unwanted QND gate vanishes and so the overall transformation becomes a simple {\small SWAP}.  Thus, somewhat remarkably, by using an incomplete transducer at most six times, one can obtain a perfect transduction operation without any need to inject input squeezing and/or perform a perfect measurement-plus-feedforward operation.

\section{Discussion}  

Our work presents both a new, useful way to classify two-mode bosonic transducers, and also shows how interference and phase-sensitive amplification can be harnessed to make a wide class of imperfect transducers perfect.  We demonstrated how this scheme could be used to accelerate state transfer into a quantum memory, even in the case where there are non-negligible levels of loss.  The approaches outlined here will help provide flexibility in designing physical tranduction systems, as it alleviates the stringent constraints on parameters that are typically needed for perfect impedance matching.   Additionally, our techniques might find applications in implementing continuous-variable logic gates \cite{Weedbrook:2012fe}.

\textbf{Acknowledgement.}  This work was supported by the AFOSR MURI FA9550-15-1-0029 on quantum transduction.

\textbf{Author contributions.}  H.-K.L. and A.A.C. contributed equally in the development of idea, derivation of results, and writing of manuscript.

\textbf{Competing interests.} The authors declare no competing interests.

\textbf{Data Availability.}  The numerical data generated in this work is available from the authors upon reasonable request.

\section{Methods}

\textbf{Local transformations.} 
The terminology and convention for Gaussian operations used in this work follows standard definitions in continuous-variable quantum information \cite{Braunstein:2005wr, Weedbrook:2012fe}.  To be self-contained, we explicitly list the relevant definitions below.  

We use the definitions $\hat{q} \equiv (\hat{a}+\hat{a}^\dag)/\sqrt{2}$, $\hat{p} \equiv (\hat{a}-\hat{a}^\dag)/\sqrt{2}i$, and $[\hat{a},\hat{a}^\dag]=1$.  The operator and quadrature transformation matrix of a phase-space rotation are given by
\begin{equation}
\hat{R}(\theta) \equiv \exp(-i\theta \hat{a}^\dag \hat{a})~~,~~
\bm{R}(\theta)= \left( \begin{array}{cc} \cos \theta & \sin \theta \\ -\sin \theta & \cos\theta \end{array} \right)~.
\end{equation}

A special rotation operator is the parity operator, i.e., $\hat{\mathcal{P}} \equiv \hat{R}(\pi)$. 
A parity operator does not mix $q$- and $p$-quadratures, but flips their sign, i.e.
\begin{equation}
\hat{\mathcal{P}}^\dag \hat{q} \hat{\mathcal{P}} = - \hat{q}~~,~~\hat{\mathcal{P}}^\dag \hat{p} \hat{\mathcal{P}} = - \hat{p}~~,~~ \bm{P} = \left( \begin{array}{cc} - 1 & 0 \\ 0 & - 1 \end{array} \right)~.
\end{equation}

The operator and quadrature transformation matrix of a single-mode squeezing operation are given by
\begin{equation}
\hat{S}(r) = \exp\left( -\frac{r}{2}(\hat{a}^2 -\hat{a}^{\dag 2}) \right)~~,~~\bm{S}(r)= \left( \begin{array}{cc} e^r & 0 \\ 0 & e^{-r} \end{array} \right)~.
\end{equation}

Our scheme requires tuning interference by modifying scattering amplitudes associated with a particular path; to do this, we consider the generalized amplifying operator $\hat{G}(\gamma)\equiv \hat{S}(r) \hat{\mathcal{P}}^n$ for $n \in \{0,1\}$, such that
\begin{eqnarray}
&\hat{G}^\dag(\gamma) \hat{q} \hat{G}(\gamma) = \gamma \hat{q}~~,~~\hat{G}^\dag(\gamma) \hat{p} \hat{G}(\gamma) = \frac{1}{\gamma} \hat{p}~,\\
&\bm{G}(\gamma)= \left( \begin{array}{cc} \gamma & 0 \\ 0 & 1/\gamma \end{array} \right)~,
\end{eqnarray}
where the amplification factor $\gamma \equiv \pm e^r$ can be any non-zero real number.  

Generally, the transformation matrix of any single mode linear transformation $\hat{U}$ must be symplectic to preserve the commutation relation, i.e.
\begin{equation}
\bm{U} \bm{\Omega}\bm{U}^\textrm{T} = \bm{\Omega}~,
~\textrm{ where }~
\bm{\Omega} \equiv \left(\begin{array}{cc} 0 & 1 \\ -1 & 0 \end{array} \right)~.
\end{equation}
Any single-mode symplectic transformation can be realized by combining squeezing and rotation operations \cite{Weedbrook:2012fe}.  Given some 2x2 full-rank real matrix $\bm{T}$, it is easy to check the following matrices are symplectic:
\begin{equation}
\bm{U} = \sqrt{\det(\bm{T})} \bm{T}^{-1}
\end{equation}
for $\det(\bm{T})>0$, and
\begin{eqnarray}
\bm{U} &=& \sqrt{|\det(\bm{T})|} \bm{T}^{-1} \bm{Z} \\
\textrm{and }~~\bm{U} &=& \sqrt{|\det(\bm{T})|} \bm{Z} \bm{T}^{-1} 
\end{eqnarray}
for $\det(\bm{T})<0$; $\bm{Z}$ is the Pauli $Z$ matrix.

\textbf{Quadrature-diagonalization and classification of two-mode linear transformation.}  
In the main text, we discussed how any two-mode linear transformation can be classified by using the rank and determinant of its transmission and reflection matrices, leading to the five classes listed in Table \ref{table:FullClass}; we also discussed the general procedure of quadrature diagonalization.  Here, we provide further technical details on both these topics.  Before doing so, we note that our classification scheme may superficially remind one of the classification of Gaussian channels \cite{Holevo:2007gn, Weedbrook:2012fe}.  We stress though that our problem is fundamentallly different:  our problem involves two-mode unitary transformations only, whereas in quantum channel theory, one is usually classifying single mode transmission in the presence of an environment (i.e.~the transformation is not unitary). 

The general procedure of quadrature-diagonalization discussed in the main text (c.f.~Eq.~(\ref{eq:diagonalize})) involves first picking a local transformation which diagonalizes one of the two output modes, i.e.
\begin{eqnarray}\label{eq:M_i}
\left( \begin{array}{c} \qout_i \\ \pout_i \end{array}\right) & =&  \bm{L}^\textrm{out}_{i} \bm{T}_{i1} \bm{L}^\textrm{in}_{1} \left( \begin{array}{c} \qin_1 \\ \pin_1 \end{array}\right) + \bm{L}^\textrm{out}_{i} \bm{T}_{i2} \bm{L}^\textrm{in}_{2} \left( \begin{array}{c} \qin_2 \\ \pin_2 \end{array}\right) \nonumber \\
&=& \left( \begin{array}{cc} \Lambda_{i1}^{(q)} & 0 \\ 0 & \Lambda_{i1}^{(p)} \end{array}\right) \left( \begin{array}{c}  \qin_1 \\ \pin_1 \end{array}\right) \nonumber \\
&&+ \left( \begin{array}{cc} \Lambda_{i2}^{(q)} & 0 \\ 0 & \Lambda_{i2}^{(p)} \end{array}\right) \left( \begin{array}{c} \qin_2 \\  \pin_2 \end{array}\right) ~.
\label{eq:MethodsDiagonal}
\end{eqnarray}
The commutation relation $[\qout_i,\pout_i] = i$  requires
\begin{equation}
\Lambda_{i1}^{(q)} \Lambda_{i1}^{(p)} + \Lambda_{i2}^{(q)} \Lambda_{i2}^{(p)}=1~.  
\end{equation}
Because local symplectic transformations do not change the determinant of transmission and reflection matrices, the above equation is equivalent to Eq.~(\ref{eq:M_det}) by recognizing
\begin{equation}
\det(\bm{T}_{i1}) = \Lambda_{i1}^{(q)} \Lambda_{i1}^{(p)} ~~,~~\det(\bm{T}_{i2}) = \Lambda_{i2}^{(q)} \Lambda_{i2}^{(p)}~.
\end{equation}

Consider next the output quadratures of the other mode $\bar{i}\neq i$.  In order to commute with output mode $i$ operators, they can only be linear combinations of the following operators:
\begin{equation}
\hat{Q}_\bot \equiv \Lambda_{i2}^{(p)} \qin_1 - \Lambda_{i1}^{(p)} \qin_2~~,~~\hat{P}_\bot \equiv \Lambda_{i2}^{(q)} \pin_1 - \Lambda_{i1}^{(q)} \pin_2~.
\end{equation}
Furthermore, because $\hat{Q}_\bot$ and $\hat{P}_\bot$ behaves as quadrature operators, i.e. $[\hat{Q}_\bot, \hat{P}_\bot]=i$, the linear combination has to be a symplectic transformation. Because $\bm{L}^\textrm{out}_{\bar{i}}$ is not involved in diagonalizing mode $i$, output mode $\bar{i}$ will have the general form
\begin{eqnarray}\label{eq:M_j}
\left( \begin{array}{c} \qout_{\bar{i}} \\ \pout_{\bar{i}} \end{array}\right) & =&  \bm{L}^\textrm{out}_{\bar{i}} \bm{T}_{\bar{i}1} \bm{L}^\textrm{in}_{1} \left( \begin{array}{c} \qin_1 \\ \pin_1 \end{array}\right) + \bm{L}^\textrm{out}_{\bar{i}} \bm{T}_{\bar{i}2} \bm{L}^\textrm{in}_{2} \left( \begin{array}{c} \qin_2 \\ \pin_2 \end{array}\right) \nonumber \\
&=& \bm{L}^\textrm{out}_{\bar{i}} \bm{U} \left( \begin{array}{cc} \Lambda_{i2}^{(p)} & 0 \\ 0 & \Lambda_{i2}^{(q)} \end{array}\right) \left( \begin{array}{c}  \qin_1 \\ \pin_1 \end{array}\right) \nonumber \\
&& + \bm{L}^\textrm{out}_{\bar{i}} \bm{U}  \left( \begin{array}{cc} -\Lambda_{i1}^{(p)} & 0 \\ 0 & -\Lambda_{i1}^{(q)} \end{array}\right) \left( \begin{array}{c} \qin_2 \\  \pin_2 \end{array}\right) ~,
\end{eqnarray}
where $\bm{U}$ is a symplectic matrix.  Because symplectic matrices are full rank, one cannot change the rank of a matrix by multiplying by them.  Therefore we have Eq.~(\ref{eq:M_eq}), i.e. 
\begin{eqnarray}
\textrm{rank}(\bm{T}_{\bar{i}1}) = &\textrm{rank}\left( \begin{array}{cc} \Lambda_{i2}^{(p)} & 0 \\ 0 & \Lambda_{i2}^{(q)} \end{array}\right) &= \textrm{rank}(\bm{T}_{i2}) , \\
\textrm{rank}(\bm{T}_{\bar{i}2}) = &\textrm{rank}\left( \begin{array}{cc} \Lambda_{i1}^{(p)} & 0 \\ 0 & \Lambda_{i1}^{(q)} \end{array}\right)& = \textrm{rank}(\bm{T}_{i1}) .
\end{eqnarray}

For classification, the output mode $\bar{i}$ in Eq.~(\ref{eq:M_j}) can be diagonalized by picking $\bm{L}^\textrm{out}_j = \bm{U}^{-1}$.  In the case where $\bar{i}=2$, and where we assume the experimentally-relevant constraint that mode $2$ cannot be squeezed, $\bm{L}^\textrm{out}_2$ is limited to being a rotation matrix, so output mode 2 cannot be quadrature-diagonalised.  Nevertheless, by using QL decomposition \cite{matrix_manual}, we can express any symplectic matrix $\bm{U}$ as
\begin{equation}\label{eq:QL}
\bm{U}=\bm{R}_U \bm{J}_U~,
\end{equation}
where $\bm{R}_U$ is a rotation matrix, $\bm{J}_U$ is lower triangular, i.e.
\begin{equation}\label{eq:J}
\bm{J}_U\equiv \left( \begin{array}{cc} \xi & 0 \\ \cdot & \cdot \end{array}\right)~.
\end{equation}
$\qout_2$ will be diagonalised if we pick $\bm{L}^\textrm{out}_2 = \bm{R}_U^{-1}$.

In Eq.~(\ref{eq:J}), the non-vanishing upper left entry, $\xi$, is the only element of this matrix that is needed for our destructive interference scheme; the remaining entries (indicated with $\cdot$ above) play no role.   Explicitly, our transduction scheme requires only the transformation matrix of the $q$-quadrature (i.e. Eq.~(\ref{eq:T_matrix})), which is sufficiently obtained from Eqs.~(\ref{eq:M_i}), (\ref{eq:M_j}), and (\ref{eq:J}).

In the following, we will discuss how Eq.~(\ref{eq:M_i}) can be constructed in each class given in Table \ref{table:FullClass}, and how each class is convertible to well known two-mode operations.  Generally, our strategy is to consider the SVD of the transmission and reflection matrices (c.f. Eq.~(\ref{eq:SVD})).

\textit{Class [[0,2]]: Identity ---}  In this class, $\bm{T}_{21}=\bm{0}$ is the null matrix.  Therefore output mode 2 involves input mode 2 quadratures only.  From Eq.~(\ref{eq:M_eq}), we have $\bm{T}_{12}=\bm{0}$, so output mode 1 involves input mode 1 quadratures only.  As a result, class [[0,2]] transformation is a tensor product of local transformation. 

Any class [[0,2]] transformation can be diagonalised by picking $\Lout_1 = \bm{T}_{11}^{-1}$, $\Lout_2 = \bm{T}_{22}^{-1}$, $\Lin_1$ and $\Lin_2$ as the identity.  The diagonalized transformation gives the relation
\begin{equation}
\qout_1 = \qin_1 ~~,~~\pout_1 = \pin_1~~,~~\qout_2 = \qin_2~~,~~\pout_2 = \pin_2 ~,
\end{equation}
which is an identity operation.

\textit{Class [[1,2]]: QND ---}  We start by diagonalising output mode 1.  The SVD of $\bm{T}_{12}$ (c.f. Eq.~(\ref{eq:SVD})) can be expressed as either
\begin{eqnarray}\label{eq:12_up}
\bm{T}_{12} &=& \bm{V}_{12} \left(\begin{array}{cc} \eta &0 \\ 0 & 0 \end{array}\right) \bm{W}_{12}~, \\
\textrm{or}~~\bm{T}_{12} &=&  \bm{V}'_{12}  \left(\begin{array}{cc} 0 &0 \\ 0 & \eta \end{array}\right) \bm{W}'_{12}~, \label{eq:12_low}
\end{eqnarray}
where $\eta \neq 0$; $ \bm{V}'_{12} \equiv \bm{V}_{12}\bm{\Omega}$ and $\bm{W}'_{12} \equiv \bm{\Omega}^\textrm{T} \bm{W}_{12}$; both $\bm{V}_{12}$, $\bm{W}_{12}$, $\bm{V}'_{12}$, and $\bm{W}'_{12}$ are rotation matrices.  

The freedom in choosing these decompositions allows us to define whether the transmitted quadrature is $\qin_2$ or $\pin_2$.  To diagonalize the transmission matrix $\bm{T}_{12}$, we can pick $\bm{L}^\textrm{out}_1=\bm{V}_{12}^{-1}$ and $\bm{L}^\textrm{in}_2=\bm{W}_{12}^{-1}$ if we want $\qin_2$ transmitted ($\Lambda^{(q)}_{12} =\eta \neq 0$ and $\Lambda^{(p)}_{12} = 0$ in Eq.~(\ref{eq:MethodsDiagonal})), or pick $\bm{L}^\textrm{out}_1=(\bm{V}'_{12})^{-1}$ and $\bm{L}^\textrm{in}_2= (\bm{W}'_{12})^{-1}$ if we want $\pin_2$ transmitted ($\Lambda^{(q)}_{12} = 0$ and $\Lambda^{(p)}_{12} =\eta \neq 0$).  Because $\det(\bm{T}_{11})=1$ in this class, $\bm{T}_{11}$ is diagonalized by choosing $\bm{L}^\textrm{in}_1 = (\bm{L}^\textrm{out}_1 \bm{T}_{11})^{-1}$.  Output mode 1 is then diagonalized, and the transformation amplitudes are given by $\Lambda_{11}^{(q)}= \Lambda_{11}^{(p)}=1$.  

For the purpose of characterisation, the output mode 2 can be diagonalised by picking $\bm{L}^\textrm{out}_2 = \bm{P} \bm{U}^{-1}$ (note that the phase-flip is added just for clarity), where $\bm{U}= \bm{T}_{22}\Lin_2 \bm{P}$.  The quadrature-diagonalised transformation is given by
\begin{eqnarray}
\qout_1 = \qin_1 + \eta \qin_2~&,&~\qout_2 = \qin_2~, \nonumber \\
\pout_1 = \pin_1~&,&~\pout_2 = -\eta\pin_1 + \pin_2 ~, \label{eq:QND1}
\end{eqnarray}
if we pick the decomposition in Eq.~(\ref{eq:12_up}), or 
\begin{eqnarray}
\qout_1 = \qin_1 ~&,&~\qout_2 = -\eta\qin_1 + \qin_2~, \nonumber \\
\pout_1 = \pin_1 + \eta \pin_2 ~&,&~\pout_2 =  \pin_2 ~, \label{eq:QND2}
\end{eqnarray}
if we pick the decomposition in Eq.~(\ref{eq:12_low}).  

Eqs.~(\ref{eq:QND1}) and (\ref{eq:QND2}) respectively represent the transformation of the QND gate $\hat{\mathcal{U}}=\exp(-i\eta \hat{p}_1 \hat{q}_2)$ and $\hat{\mathcal{U}}=\exp(i\eta \hat{q}_1 \hat{p}_2)$.  These two QND gates are equivalent by interchanging the $q$- and $p$-quadratures in both modes.  It is thus clear that the decompositions in Eqs.~(\ref{eq:12_up}) and Eq.~(\ref{eq:12_low}) are equivalent up to simple local rotations.

We note that for any two-mode transformation in this class, the effective QND strength $\eta$ is {\it not} invariant under local transformation.  Explicitly, if we choose $\bm{L}^\textrm{out}_1 = \bm{G}(\gamma)\bm{V}_{12}^{-1}$, the transformation still has the diagonal form of a QND gate, but the QND strength is modified as $\eta \rightarrow \gamma \eta$.

Recall that we are interested in situations where it is not possible to implement squeezing operations on mode 2.  Both output $q$-quadratures can still be diagonalized by using the QL decomposition as discussed before.  The $q$-quadrature transformation matrix Eq.~(\ref{eq:T_matrix}) is
\begin{equation}
\bm{T}^{qq} =  \left( \begin{array}{cc} 1 & \eta \\ 0 & -\xi  \end{array}\right)~~\textrm{or}~~\bm{T}^{qq} =  \left( \begin{array}{cc} 1 & 0 \\ \xi \eta  & -\xi  \end{array}\right),
\end{equation}
if the choice of decomposition is Eq.~(\ref{eq:12_up}) or Eq.~(\ref{eq:12_low}) respectively.  The freedom to choose between these $\bm{T}^{qq}$ guarantees that one can always arrange the amplitude of the transmission-transmission path in  Eq.~(\ref{eq:interference}) to be non-vanishing.

\textit{Classes $[[2,2]]$ ---}  Depending on the sign of determinant of the transmission matrix (and correspondingly that of reflection matrix according to Eq.~(\ref{eq:M_det})), different choice of local operation is required to diagonalize the transformation.  For clarity, we separately discuss the three cases: $0>\det(\bm{T}_{21})$, $1>\det(\bm{T}_{21})>0$, and $\det(\bm{T}_{21})>1$.  Interestingly, the equivalent well-known transformation in each case would be different.  We note that all transformations in these three classes can be diagonalized while still obeying the practical constraint on mode $2$, i.e.~mode 2 cannot be squeezed.  Besides, all elements in the $q$-quadrature transformation matrix Eq.~(\ref{eq:T_matrix}) is non-vanishing in this class.

\textit{$0>\det(\bm{T}_{21})$: TMS ---}
Output mode 2 can be quadrature-diagonalized by taking  $\bm{L}^\textrm{in}_{2}=\bm{W}_{22}^{-1}$, $\bm{L}^\textrm{out}_2 = \bm{V}_{22}^{-1}$, and $\bm{L}^\textrm{in}_1 = \sqrt{|\det(\bm{T}_{21})|} \bm{Z} (\bm{L}^\textrm{out}_2 \bm{T}_{21})^{-1}$
The coefficients in the diagonalized transformation (c.f.~Eq.~(\ref{eq:MethodsDiagonal})) are given by
\begin{eqnarray}
&\Lambda_{21}^{(q)}=-\Lambda_{21}^{(p)} = \sqrt{|\det(\bm{T}_{21})|} \equiv \sinh r~, \\
&\Lambda_{22}^{(q)}=D_{22}^{(q)}~~,~~\Lambda_{22}^{(p)}=D_{22}^{(p)}~,
\end{eqnarray}
where
\begin{equation}
\sqrt{D_{22}^{(q)}D_{22}^{(p)}} =  \sqrt{\det(\bm{T}_{22})} \equiv \cosh r~.
\end{equation}

Next, output mode 1 can be diagonalised by picking $\bm{L}^\textrm{out}_1 = \bm{U}^{-1}$, where $\bm{U}= (\bm{T}_{12} \Lin_2 \bm{Z})/\sinh r$.  The $q$-quadrature transformation matrix is then
\begin{equation}
\bm{T}^{qq} =  \left( \begin{array}{cc} D^{(p)}_{22} & \sinh r \\ \sinh r  & D^{(q)}_{22}  \end{array}\right)~.
\end{equation}

For the purpose of characterization, we can pick the same local operations except $\Lin_2=(\bm{D}_{22}\bm{W}_{22})^{-1}$.  The fully diagonalized transformation becomes
\begin{eqnarray} \label{eq:TMS}
\qout_1 &=& \cosh r \qin_1 + \sinh r \qin_2 ~, \nonumber \\
\pout_1 &=& \cosh r \pin_1 - \sinh r \pin_2 ~, \nonumber \\
\qout_2 &=& \sinh r \qin_1 + \cosh r \qin_2 ~, \nonumber \\
\pout_2 &=& -\sinh r \pin_1 + \cosh r \pin_2~.
\end{eqnarray}
This is the transformation of a two-mode-squeezing operation, i.e. $\hat{\mathcal{U}}=\exp\left(-i r (\hat{q}_1 \hat{p}_2 + \hat{p}_1 \hat{q}_2) \right)$.  Here the TMS strength $r$ is fixed by the determinant of transmission and reflection matrices, and cannot be altered by local operations.

\textit{$1>\det(\bm{T}_{21})>0$: BS ---}  
Output mode 2 can be quadrature-diagonalised by picking  $\bm{L}^\textrm{in}_{2}=\bm{W}_{22}^{-1}$, $\bm{L}^\textrm{out}_2 = \bm{V}_{22}^{-1}$, and $\bm{L}^\textrm{in}_1 = \sqrt{\det(\bm{T}_{21})} \bm{P} (\bm{L}^\textrm{out}_2 \bm{T}_{21})^{-1}$.  The coefficients in the diagonalized form (c.f.~Eq.~(\ref{eq:MethodsDiagonal})) are then
\begin{eqnarray}
& \Lambda_{21}^{(q)}=\Lambda_{21}^{(p)} = -\sqrt{\det(\bm{T}_{21})} \equiv -\sin \theta~, \\
& \Lambda_{22}^{(q)}=D_{22}^{(q)}~~,~~\Lambda_{22}^{(p)}=D_{22}^{(p)}~.
\end{eqnarray}
We note that
\begin{equation}
\sqrt{D_{22}^{(q)}D_{22}^{(p)}} = \sqrt{\det(\bm{T}_{22})} \equiv \cos \theta~.
\end{equation}

Output mode 1 can be diagonalized by picking $\bm{L}^\textrm{out}_1 = \bm{U}^{-1}$, where $\bm{U}= (\bm{T}_{12} \Lin_2)/\sin\theta$.  
The $q$-quadrature transformation matrix  is then
\begin{equation}
\bm{T}^{qq} =  \left( \begin{array}{cc} D^{(p)}_{22} & \sin\theta \\ -\sin\theta  & D^{(q)}_{22}  \end{array}\right)~.
\end{equation}

For the purpose of characterization, we can choose the same local transformations except $\Lin_2=(\bm{D}_{22}\bm{W}_{22})^{-1}$.  The final diagonalized quadrature transformation becomes
\begin{eqnarray}
\qout_1 = \cos\theta \qin_1 + \sin \theta \qin_2~&,&~\qout_2 = -\sin\theta \qin_1 + \cos \theta \qin_2~, \nonumber \\
\pout_1 = \cos\theta \pin_1 + \sin \theta \pin_2~&,&~\pout_2 = -\sin\theta \pin_1 + \cos \theta \pin_2~. \nonumber \\
\end{eqnarray}
This is the transformation of a BS, i.e.~$\hat{\mathcal{U}}= \exp\left(i\theta (\hat{q}_1 \hat{p}_2- \hat{p}_1 \hat{q}_2) \right)$.  We note that the BS angle $\theta$ is fixed by the determinant of the transmission and reflection matrices, and thus cannot be altered by local operations.


We pause to note an interesting physical consequence of our discussion here and classification scheme: TMS and BS operations {\it cannot} be made equivalent using purely local operations,
because their determinant of transmission matrix are different.  At first glance, this is surprising, as it is well known that both these operations can produce two-mode squeezed vacuum states with appropriately prepared separable input states.  In particular, one can either apply TMS on the vacuum state of two modes, or by locally squeezing the vacuum of two modes before sending them through a BS.  Our result shows that this correspondence is not general, but only holds for a particular choice of input states.

\textit{$\det(\bm{T}_{21})>1$: swapped TMS ---}
Output mode 2 can be quadrature-diagonalized by taking $\bm{L}^\textrm{in}_{2}=\bm{W}_{22}^{-1}$, $\bm{L}^\textrm{out}_2 = \bm{V}_{22}^{-1}$, and $\bm{L}^\textrm{in}_1 = \sqrt{\det(\bm{T}_{21})} (\bm{L}^\textrm{out}_2 \bm{T}_{21})^{-1}$
The diagonal form (c.f.~Eq.(\ref{eq:MethodsDiagonal})) is then determined by:
\begin{eqnarray}
&\Lambda_{21}^{(q)}=\Lambda_{21}^{(p)} = \sqrt{\det(\bm{T}_{21})} \equiv \cosh r~, \\
&\Lambda_{22}^{(q)}=D_{22}^{(q)}~~,~~\Lambda_{22}^{(p)}=D_{22}^{(p)}~,
\end{eqnarray}
where
\begin{equation}
\sqrt{-D_{22}^{(q)}D_{22}^{(p)}} =  \sqrt{|\det(\bm{T}_{22})|} \equiv \sinh r~.
\end{equation}

Next, output mode 1 can be diagonalised by picking $\bm{L}^\textrm{out}_1 = \bm{U}^{-1}$, where $\bm{U}= (\bm{T}_{12} \Lin_2 \bm{P})/\cosh r$.  The $q$-quadrature transformation matrix is then
\begin{equation}
\bm{T}^{qq} =  \left( \begin{array}{cc} -D^{(p)}_{22} & \cosh r \\ \cosh r  & D^{(q)}_{22}  \end{array}\right)~.
\end{equation}

For classification purpose, we take the same local operations except $\Lin_2=(\bm{Z}\bm{D}_{22}\bm{W}_{22})^{-1}$.  The quadrature transformation is then
\begin{eqnarray}
\qout_1 &=& \sinh r \qin_1 + \cosh r \qin_2 ~, \nonumber \\
\pout_1 &=& -\sinh r \pin_1 + \cosh r \pin_2 ~, \nonumber \\
\qout_2 &=& \cosh r \qin_1 + \sinh r \qin_2 ~, \nonumber \\
\pout_2 &=& \cosh r \pin_1 - \sinh r \pin_2~.
\end{eqnarray}
This transformation is equivalent to a composition of a {\small SWAP} and a TMS, i.e.~$\hat{\mathcal{U}}=\exp\left(-i r (\hat{q}_1 \hat{p}_2 + \hat{p}_1 \hat{q}_2) \right)\hat{\mathbb{S}}$.  The TMS strength $r$ is similarly fixed by the determinant of transmission and reflection matrices, and cannot be altered by local operations.

We refer this operation (product of SWAP and TMS) as a \textit{swapped TMS} operation; it is a less discussed class of two-mode transformation.  While both swapped TMS and TMS generate nonlocal excitations, they belong to different classes and so they cannot be made equivalent using local operations only.  Further, although it involves a {\small SWAP} operation, a swapped TMS operation cannot be employed directly in transduction,  because the unwanted TMS part of the transformation cannot be eliminated via local operations only.


\textit{Class [[2,1]]: swapped QND ---}
In this class, we start by diagonalizing output mode 2.  We first consider the SVD of $\bm{T}_{22}$,
\begin{equation}\label{eq:class21_M22}
\bm{T}_{22}=\bm{V}_{22} \left( \begin{array}{cc} 0 & 0 \\ 0 & \eta \end{array} \right) \bm{W}_{22}~.
\end{equation}
$\bm{T}_{22}$ is diagonalised by picking $\bm{L}^\textrm{out}_2=\bm{V}_{22}^{-1}$ and $\bm{L}^\textrm{in}_2 = \bm{W}_{22}^{-1}$.   Next, $\bm{T}_{12}$ is diagonalised by picking $\bm{L}^\textrm{in}_1 = (\bm{L}^\textrm{out}_2 \bm{T}_{21})^{-1}$.  These processes yield a diagonal form (c.f.~Eq.~(\ref{eq:MethodsDiagonal})) determined by $\Lambda^{(q)}_{21}=\Lambda^{(p)}_{21}=1$, $\Lambda^{(q)}_{22}=0$, and $\Lambda^{(q)}_{22}=\eta$.  

The transformation of output mode 1 is given by Eq.~(\ref{eq:M_j}), so it can be diagonalized by picking $\bm{L}^\textrm{out}_{1}= \bm{P}\bm{U}^{-1}$.  The diagonalized quadrature transformation is given by 
\begin{eqnarray}
\qout_1 = -\eta \qin_1 + \qin_2 ~&,&~\qout_2 = \qin_1~,   \\
\pout_1 = \pin_2~&,&~\pout_2 = \pin_1 + \eta \pin_2 ~.  \label{eq:SQND_2}
\end{eqnarray}
As discussed in main text, this is the transformation of a sQND (c.f. Eq.~(\ref{eq:SQND_gate})).

We note that the decomposition in Eq.~(\ref{eq:class21_M22}) is not unique: we can choose a non-vanishing upper left entry by using $\Lout_2=(\bm{\Omega}\bm{V}_{22})^{-1}$ and $\bm{L}^\textrm{in}_2 = (\bm{W}_{22}\bm{\Omega})^{-1}$.  The subsequent procedure of diagonalisation is similar, and the diagonalised transformation remains a swapped QND gate except the $q$- and $p$-quadratures are interchanged.

\textit{Class [[2,0]]: {\small SWAP} ---}
In this class $\bm{T}_{22}=\bm{T}_{11}=\bm{0}$, so the output mode only contains quadratures from the opposite input mode, and hence the transformation is a {\small SWAP} up to local transformation.  Explicitly, the transformation can be diagonalized by picking $\bm{L}^\textrm{in}_1 = \bm{T}_{21}^{-1}$, $\bm{L}^\textrm{out}_1 = \bm{T}_{12}^{-1}$, while both $\bm{L}^\textrm{in}_2$ and $\bm{L}^\textrm{out}_2$ are taken to be the identity.   We then have
\begin{equation}
\qout_1 = \qin_2 ~~,~~\pout_1 = \pin_2~~,~~\qout_2 = \qin_1~~,~~\pout_2 = \pin_1 ~,
\end{equation}
which indicates a {\small SWAP}.

\textbf{Complete transduction with two TMS.}
In the main text, we gave an example for how our interference-based scheme could implement perfect transduction by using two incomplete BS operations.  Here we analyze a more counter-intuitive example:  complete transduction by using two sequential TMS operations.  TMS is usually viewed as a process that generates correlated excitations, and is thus not directly related to or suited for state transfer.  Nonetheless, our scheme allows TMS operations to be exploited for perfect state transfer, as we now show.

For simplicity, we assume both TMS operations are identical,  and that each transforms quadratures as given in Eq.~(\ref{eq:TMS}).  With the arrangement in Fig.~\ref{fig:interfere}, we can see that the amplitude associated with two consecutive $q$-quadrature reflections is $\cosh^2 r $ (with $r$ the squeezing strength of each TMS).  Similarly, the amplitude associated with two consecutive transmissions is $\gamma \sinh^2 r$.  These two paths interfere destructively when $\gamma = -\coth^2 r $.  The overall transformation is then given by
\begin{eqnarray}
&\qout_1 = -\eta \coth^2 r \qin_1 -\coth r \qin_2~~,~~\qout_2 = - \coth r \qin_1 ~, \nonumber \\
&\pout_1 = -\tanh r \pin_2~~,~~ \pout_2 = -\tanh r \pin_1 +\eta \pin_2~,
\end{eqnarray}
where $\eta = 1 + \tanh^2 r$.

If now appropriate local operations are applied before and after the concatenated TMS, i.e.~$\Lin_1 =\Lout_1= \bm{G}(-\tanh r)$, the overall transformation becomes the standard form of a sQND in Eq.~(\ref{eq:SQND_quad}).  One-way transduction can be completed by injecting infinitely squeezed state or homodyne detection, as described in main text.

\textbf{Imperfect measurement/squeezing and noise-tolerant bosonic codes.}  Here we discuss the detrimental effect on transduction when imperfect input squeezing and measurement is employed with sQND operation.
Without loss of generality, 
we consider a transduction from mode 1 to 2.  The two-mode input state is $|\Psi^\textrm{in}\rangle \equiv |\psi_0\rangle |\psi_\textrm{anc}\rangle$, where $|\psi_0 \rangle$ is the mode-1 input state that we wish to transfer, and $|\psi_\textrm{anc} \rangle$ is the auxiliary state that is prepared in mode 2.  After sQND, the output state becomes
\begin{equation}
|\Psi^\textrm{out}\rangle \equiv \hat{\mathcal{U}}_\textrm{sQ} |\Psi^\textrm{in}\rangle = e^{i \eta \hat{p}_1 \hat{q}_2} |\psi_\textrm{anc}\rangle |\psi_0\rangle~.
\end{equation}

We first consider the case where input squeezing is used to mitigate the unwanted QND interaction.  The ancilla state in this case would ideally be an infinite squeezed vacuum; we consider the more realistic case of a finitely squeezed vacuum, $|\psi_\textrm{anc}\rangle = \hat{S}(r)|\textrm{vac}\rangle$.  If we trace out mode 1 in the final output state, the resulting state of output mode 2 is given by
\begin{eqnarray}
\rho^\textrm{out} &=& \Tro{|\Psi^\textrm{out}\rangle \langle \Psi^\textrm{out}|} \nonumber \\
&=& \int dp \frac{1}{\sigma\sqrt{\pi}} e^{-\left(\frac{p}{\sigma} \right)^2} e^{i p \hat{q}_2} |\psi_0\rangle \langle \psi_0 | e^{-i p \hat{q}_2}~,  \label{eq:squeeze_imperfect}
\end{eqnarray}
where $\sigma = \eta e^{-r}$.  We thus see that the output state is a ensemble of displaced versions of the input state $| \psi_0 \rangle$, where the displacement is of the $\hat{p}$ quadrature and follows a Gaussian distribution.  The width of the Gaussian distribution depends on the degree of squeezing.  It is easy to see that when squeezing is infinite, i.e. $r\rightarrow \infty$, the Gaussian distribution becomes a Dirac delta function and so the transduction is perfect, i.e. $\rho^\textrm{out}=|\psi_0\rangle \langle \psi_0 |$.

For the complementary approach where homodyne detection plus feedforward is used to undo the unwanted QND interaction, we first study the ideal case where the measurement is perfect.  For simplicity, we assume the auxiliary state is vacuum, i.e. $\ket{\psi_\textrm{anc}}=\ket{\vac}$.  Homodyne measurement of output mode 1 will result in a measurement outcome $p_1$; this measurement result is then used to perform a displacement of the mode-2 $p$-quadrature.  This result is a displacement of the mode-2 state that is conditioned on the measurement outcome.  Averaging over possible measurement outcomes, the ensemble-averaged output state after measurement plus feedforward becomes
\begin{eqnarray}
\rho^\textrm{out} &=& \int dp_1 \Tro{|p_1\rangle \langle p_1| e^{-i\eta_D p_1 \hat{q}_2}|\Psi^\textrm{out}\rangle \langle \Psi^\textrm{out}|e^{i\eta_D p_1 \hat{q}_2}} \nonumber \\
&=& \Tro{e^{-i\eta_D \hat{p}_1 \hat{q}_2}|\Psi^\textrm{out}\rangle \langle \Psi^\textrm{out}|e^{i\eta_D \hat{p}_1 \hat{q}_2}} \nonumber \\
&=& \int dp \frac{1}{(\eta -\eta_D) \sqrt{\pi}} e^{-\left(\frac{p}{\eta -\eta_D} \right)^2} e^{i p \hat{q}_2} |\psi_0\rangle \langle \psi_0 | e^{-i p \hat{q}_2}~, \nonumber \\
\label{eq:homo_perfect}
\end{eqnarray}
where $\eta_D$ corresponds to strength of the feedforward operation (i.e.~it is the proportionality constant between the measurement outcome and the applied displacement).  In the second step above, we employed the fact that $\hat{p}|p_1\rangle = p_1 |p_1\rangle$.  

The transduction is complete (i.e.~perfect) when we choose  $\eta_D =  \eta$.  Physically it means the unwanted QND gate is exactly cancelled by the measurement plus feedforward operation;  mathematically, in Eq.~(\ref{eq:homo_perfect}) the Gaussian distribution of displacements becomes a Dirac delta function at $p=0$, so there is no displacement noise.

Having understood the perfect measurement case, we now turn to an imperfect (inefficient) measurement.  This can be modelled by performing a BS operation between mode one and an environment mode before homodyne detection is done.  For a detection inefficiency $\epsilon$, the total output state before homodyne detection is
\begin{equation}
|\Psi^\textrm{HD}\rangle \equiv e^{i \eta (\sqrt{1-\epsilon} \hat{p}_1+\sqrt{\epsilon}\hat{p}_E)\hat{q}_2} |\vac\rangle |\psi_0\rangle |\vac_E\rangle~,
\end{equation}
where the subscript $E$ denotes the environment mode (associated with the other BS port); we assume vacuum noise for simplicity.  After homodyne detection and the corresponding feedforward displacement, the output state in mode 2 becomes
\begin{eqnarray}
\rho^\textrm{out} &=& \int dp_1 \Troe{|p_1\rangle \langle p_1| e^{-i\eta_D p_1 \hat{q}_2}|\Psi^\textrm{HD}\rangle \langle \Psi^\textrm{HD}|e^{i\eta_D p_1 \hat{q}_2}} \nonumber \\
&=& \int dp \frac{1}{\sigma \sqrt{\pi}} e^{-\left(\frac{p}{\sigma} \right)^2} e^{i p \hat{q}_2} |\psi_0\rangle \langle \psi_0 | e^{-i p \hat{q}_2}~, \label{eq:homo_imperfect}
\end{eqnarray}
where $\sigma \equiv \sqrt{\eta^2 +\eta_D^2 -2\eta \eta_D \sqrt{1-\epsilon}}$.  The variance of random displacement is minimised when we choose $\eta_D = \eta \sqrt{1-\epsilon} $, which gives $\sigma = \eta \sqrt{\epsilon}  $.  

As seen from Eqs.~(\ref{eq:squeeze_imperfect}) and (\ref{eq:homo_imperfect}), both finite injected squeezing and inefficient homodyne detection induce the same detrimental effect: the transmitted state is corrupted by a displacement noise in the $p$-quadrature.  
One can now ask whether this kind of corrupted transduction is of any practical utility. The answer, not surprisingly, is dependent on the kind of state one is trying to transduce.  Consider the very relevant example where we are interested in transferring a logical qubit state which is encoded in the initial mode-1 state.  In this case, the fidelity of the transmitted logical qubit is closely related to the type of bosonic code employed \add{and the decoding/recovery procedures} \cite{2018PhRvA..97c2346A}.  

Here we give specific examples of how the choice of bosonic code can be exploited to better preserve the qubit state after an imperfect transduction.  The mode-1 input state encodes a qubit, and thus has the general form:
\begin{equation}\label{eq:thetaphi_input}
|\psi_0\rangle = |\vartheta, \varphi\rangle \equiv \cos \frac{\vartheta}{2}|0_L\rangle + e^{i\varphi} \sin\frac{\vartheta}{2} |1_L\rangle~,
\end{equation}
where $\vartheta$ and $\varphi$ parametrize the qubit information, and $|0_L\rangle$,$|1_L\rangle$ are the basis states in the chosen bosonic code.  \add{In view of the difficulties of QND measurement and full control in some bosonic platforms (e.g. spin ensembles and optical modes), here we focus on the passive error tolerance of the bosonic code without any decoding and recovery processes.
}
For any pure-state encoding, i.e. $|0_L\rangle$ and $|1_L\rangle$ are pure states, the logical fidelity of the output state is then given by the physical fidelity between the input and output state, i.e.
\begin{eqnarray}\label{eq:normal_code}
\mathcal{F}(\vartheta, \varphi) &\equiv& 
\langle \psi_0 | \rho^\textrm{out}( \vartheta,\varphi) | \psi_0 \rangle \nonumber \\
& = & \int dp \frac{1}{\sigma \sqrt{\pi}} e^{-\left(\frac{p}{\sigma} \right)^2} \left| \langle  \vartheta,\varphi | e^{i p \hat{q}_2} | \vartheta,\varphi\rangle \right|^2~.
\end{eqnarray}
The performance of our qubit transduction can be characterized by the average qubit fidelity
\begin{equation}\label{eq:average_fidelity}
\bar{\mathcal{F}} \equiv \frac{1}{4\pi} \int_0^{2\pi} \int_0^\pi\mathcal{F}(\vartheta, \varphi) \sin\theta d\vartheta d\varphi ~.
\end{equation}

\begin{figure}
\begin{center}
\includegraphics[width=\linewidth]{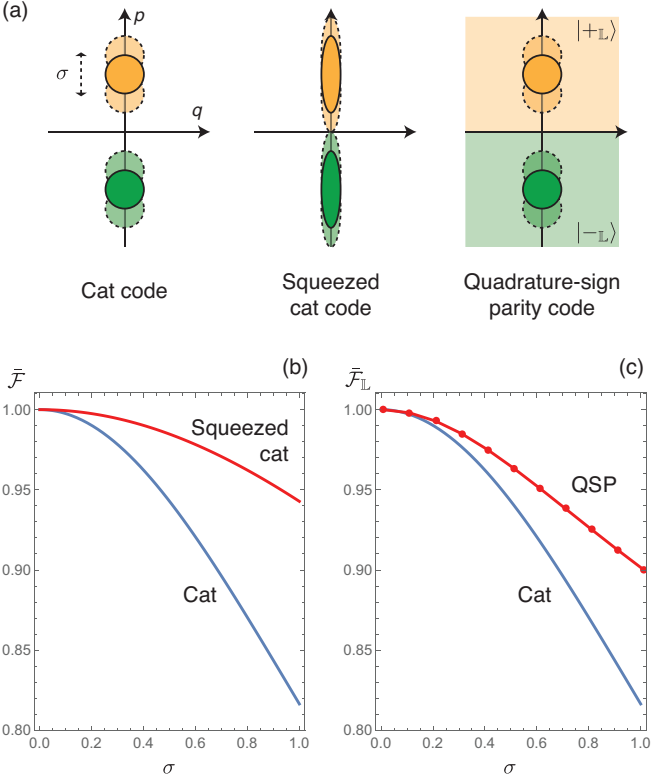}
\caption{{\bf Error tolerance of bosonic codes.}
(a) Schematic of various single-mode bosonic codes of a logical qubit.  Solid orange (green) eclipse denotes the pure-state basis states $|+_L\rangle$ ($|-_L\rangle$) of each encoding.  Dotted eclipses denote the basis state after random displacement in the $p$-quadrature.  The shaded area for the mixed-state QSP encoding denotes the region within which a physical state can represent the mixed-state logical basis.
(b) State-averaged logical qubit fidelity as a function of the variance $\sigma$ of $p$-quadrature displacement noise.  The cat code has displacement $\alpha=2$ (blue), and the squeezed cat code has $\alpha=2$ and $e^r = 1/2$ (red).
(c) State-averaged logical qubit fidelity versus displacement noise for cat code state with $\alpha=2$ (blue),  and for the case where the final state is treated as a QSP encoded qubit (red with dots).  \add{We note that these fidelities quantify the uncorrected noise tolerance of the bosonic codes, but not their ultimate performance under optimal decoding and recovery.}
\label{fig:fidelity}}
\end{center}
\end{figure}

For any encoding \add{bases} $\{|0_L\rangle, |1_L\rangle \}$, we find that the logical fidelity can be improved by simply using squeezed versions of the original encoding states  $\{|0_S\rangle \equiv \hat{S}|0_L\rangle, |1_S\rangle \equiv\hat{S}|1_L\rangle \}$, i.e.  $|\psi_0\rangle =\hat{S} |\vartheta, \varphi\rangle$.
The average fidelity of the squeezed encoding is readily computed as
\begin{equation}\label{eq:squeeze_code}
\mathcal{F}_S(\vartheta, \varphi)= \int dp \frac{1}{\sigma e^{r} \sqrt{\pi}} e^{-\left(\frac{p}{\sigma e^{r}} \right)^2} \left| \langle  \vartheta,\varphi | e^{i p \hat{q}_2} | \vartheta,\varphi\rangle \right|^2 ~. 
\end{equation}
When comparing with Eq.~(\ref{eq:normal_code}), the squeezed encoding effectively modifies the variance of the random displacement distribution, i.e. $\sigma \rightarrow \sigma e^r$.  The variance is reduced if $r < 0$, i.e. the squeezing $\hat{S}$ is a $p$-quadrature amplification.  The intuition behind this strategy is simple: for a fixed amount of $p$-quadrature noise, its relative significance would be reduced if the information encoded in $p$-quadrature is amplified.  In Fig.~\ref{fig:fidelity}, we show an explicit numerical result for the cat code \cite{Ralph:2003jk}, i.e. 
\begin{equation}\label{eq:cat_code}
|0/1_L\rangle = \frac{1}{\mathcal{N}_\pm} \left(|i \alpha\rangle \pm |-i \alpha\rangle \right)~,
\end{equation}
where $|\pm i \alpha\rangle$ is a coherent state with purely imaginary amplitude $\pm i \alpha$; $\mathcal{N}_\pm \equiv \sqrt{2 (1\pm e^{-2|\alpha|^2})}$.  \add{We note that appropriately squeezing a bosonic state can also improve the tolerance of encoded quantum information against channel loss \cite{2013PhRvA..87d2308F, LeJeannic:2018ek}.}

Apart from choosing another encoding, the logical fidelity can also be improved by using noiseless subsystems (NS) \cite{Knill:2000wt, Zanardi:2001vo, Kempe:2001bg}.  The idea of NS is to encode a qubit information not by two states, but by any state in two subspaces.  A consequence is that if an encoding state is corrupted by noise, the logical information is not lost if the corrupted state remains in the same subspace.

In bosonic systems, adapting NS allows a pure logical state to be represented by a mixed physical state \cite{2017PhRvA..95b2303L}.  To illustrate how NS can enhance the noise tolerance, we consider a variant of the recently proposed quadrature-sign parity (QSP) encoding \cite{Marshall2018}, which represents the logical computational basis by the parity of the bosonic state, and the qubit coherence by the sign of the quadrature of the bosonic wave function.  It has been shown that the basis states of cat code lie within the encoding subspace of QSP encoding.

The intuition of improved noise tolerance could be understood from Fig.~\ref{fig:fidelity}a.  Under displacement fluctuation, each coherent state component of a cat state roughly remains in the same side of the phase space, i.e. its wavefunction has the same sign of $p$-quadrature.  Therefore, although the encoding state is transformed by displacement noise, the logical information is retained.  Qualitatively, we compute the QSP logical fidelity for a noisy cat code qubit as \cite{Marshall2018}
\begin{eqnarray}
\mathcal{F}_\mathbb{L}(\vartheta, \varphi) &=& \frac{1}{2}\Big(1+ \sin\theta \cos\phi\langle \hat{X}_M\rangle  \nonumber\\
&&+ \sin\theta \sin\phi \langle \hat{Y}_M\rangle + \cos\theta  \langle \hat{Z}_M\rangle  \Big)~,
\end{eqnarray}
where $\langle \hat{O}\rangle \equiv \Tr{\hat{O}\rho}$; the logical operators are
\begin{equation}\label{eq:QSP}
\hat{X}_M = \int dp \Theta(p) |p\rangle\langle p|  ~;~\hat{Z}_M = e^{i\pi \hat{a}^\dag \hat{a}}~;~\hat{Y}_M \equiv i \hat{X}_M \hat{Z}_M~;
\end{equation}
$\Theta(p)$ is the sign function.  The average fidelity is still obtained by Eq.~(\ref{eq:average_fidelity}).  As shown in Fig.~\ref{fig:fidelity}c, QSP encoding can improve the logical fidelity of a noisy cat-code qubit.  

In summary, even with finite injected squeezing and imperfect measurement, our scheme produces a very specific kind of noise (random displacement noise in one quadrature only); this allows us to develop noise-tolerant bosonic codes of qubit states.  \add{We note that the logical fidelity can be further improved by conducting active error-correction \cite{2018PhRvA..97c2346A}}; this kind of single-quadrature noise also allows simpler error-correction protocols (c.f.  supplementary information in Ref.~\cite{Higginbotham:2018ca}).

\textbf{Perfect two-way transduction via interference }
Here, we provide more details on the procedure presented in Eq.~(\ref{eq:6pass}) of the main text, where interference is used to accomplish perfect transduction without any need for injected squeezing or a homodyne measurement.  The first and the last swapped QND gates, $\hat{\mathcal{U}}_\textrm{sQ}(\eta_1)$ and $\hat{\mathcal{U}}_\textrm{sQ}(\eta_3)$ can be realised by the procedure described in main text.  The second swapped QND gate, $\hat{\mathcal{U}}_\textrm{sQ}^\dag(\eta_2)$, can be realised by implementing $\hat{\mathcal{U}}_\textrm{sQ}(\eta_2)$, and applying local rotations before and afterwards, i.e.
\begin{equation}
\hat{\mathcal{U}}^\dag_\textrm{sQ}(\eta_2) = \hat{R}_1(-\frac{\pi}{2})\hat{R}_2(-\frac{\pi}{2}) \hat{\mathcal{U}}_\textrm{sQ}(\eta_2)\hat{R}_1(\frac{\pi}{2})\hat{R}_2(\frac{\pi}{2})~,
\end{equation}
where $\hat{R}_l$ denotes the rotation at mode $l$.

In the circuit diagram Fig.~\ref{fig:6pass}, we have derived a circuit (Fig.~\ref{fig:6pass}b) that is equivalent to the original transformation in Eq.~(\ref{eq:6pass}).  Our strategy is to sandwich each QND gate by an amplification ($\hat{G}$) and a deamplification ($\hat{G}^\dag$) of the same strength.  This arrangement will modify the QND gate strength as \cite{H_amp}
\begin{eqnarray}
&&\hat{G}_1^\dag(\gamma) \exp(i\eta \hat{p}_1 \hat{q}_2) \hat{G}_1(\gamma) = \exp(i  \frac{\eta}{\gamma} \hat{p}_1 \hat{q}_2)~, \nonumber \\
&=&\hat{G}_2(\gamma) \exp(i\eta \hat{p}_1 \hat{q}_2) \hat{G}^\dag_2(\gamma) ~,
\end{eqnarray}
where the subscript of $\hat{G}$ denotes the mode it is applying on.  

The three QND gates can be merged to be a single QND gate with strength 
\begin{equation}\label{eq:3QND}
\frac{1}{\gamma_1}\eta_3 +\frac{1}{\gamma_1 \gamma_2} (-\eta_2) + \frac{1}{\gamma_2}\eta_1 = \frac{\gamma_2 \eta_3 - \eta_2 + \gamma_1 \eta_1}{\gamma_1 \gamma_2}~.
\end{equation}
This strength vanishes if local amplification $\gamma_1$ and $\gamma_2$ are chosen to satisfy $\eta_2= \gamma_1 \eta_1 + \gamma_2 \eta_3 $, which is always possible for any non-vanishing $\eta_1$, $\eta_2$, $\eta_3$.  This will eliminate the QND gate, and so the circuit in Fig.~\ref{fig:6pass}c is reduced to a perfect {\small SWAP}.

\textbf{Noise in lossy beam splitter.}
 We here present the analysis of the performance of our scheme when one of the mode is lossy.  This situation can describe a practical application where a quantum state is transferred between a lossy quantum processor and an essentially lossless quantum memory.  By directly integrating Eq.~(\ref{eq:LBS}), the solution for the final-time mode operators is given by
\begin{equation}\label{eq:LBS_1}
\left( \begin{array}{c} \hat{a}_1(t) \\ \hat{a}_2 (t) \end{array} \right) =
\left( \begin{array}{cc} \tilde{T}_{11} & \tilde{T}_{12} \\ \tilde{T}_{21} & \tilde{T}_{22} \end{array} \right) \left( \begin{array}{c} \hat{a}_1(0) \\ \hat{a}_2 (0) \end{array} \right) 
+ \sum_{k\geq 3} \left( \begin{array}{c} \tilde{T}_{1k}\Ain_k \\ \tilde{T}_{2k} \Ain_k \end{array} \right)~.
\end{equation}
The environmental noise is injected through the temporal bath modes:
\begin{equation}
\Ain_k \equiv \int_0^t \mathcal{A}_k(t') \hat{B}^\textrm{in}(t') dt'~,
\end{equation}
where $ \hat{B}^\textrm{in}(t) $ is the bath operator that satisfies $[ \hat{B}^\textrm{in}(t),  \hat{B}^{\textrm{in}\dag}(t')]=\delta(t-t') $ and $[ \hat{B}^\textrm{in}(t),  \hat{B}^{\textrm{in}}(t')]=0 $.
$\mathcal{A}_k(t)$ are functions of $t$ that describe the temporal profile of the bath modes; they are orthogonal such that each temporal bath mode behaves as an independent bosonic mode, i.e. $[\Ain_k, \Aind_l ]= \delta_{kl}$ and $[\Ain_k, \Ain_l ]= 0$.  We use $k=1,2$ to denote the system source and target modes as always, and use $k\geq 3$ to label additional temporal modes of the environment.   

The transformation matrix of the system modes can be obtained analytically as
\begin{eqnarray}\label{eq:T_tilde}
&&\left( \begin{array}{cc} \tilde{T}_{11} & \tilde{T}_{12} \\ \tilde{T}_{21} & \tilde{T}_{22} \end{array} \right) \nonumber \\
&=& \frac{e^{-\theta \sin \Gamma}}{\cos \Gamma} 
\left( \begin{array}{cc} \cos\big((\theta \cos \Gamma) + \Gamma \big) & \sin(\theta \cos \Gamma)  \\ -\sin(\theta \cos \Gamma) & \cos\big((\theta \cos \Gamma) - \Gamma \big) \end{array} \right)~. \nonumber \\
\end{eqnarray}
The coherent and dissipative BS angles, $\theta$ and $\Gamma$, are respectively
\begin{equation}
\theta \equiv g t~~,~~e^{i\Gamma} \equiv \sqrt{1-\left(\frac{\kappa}{4g} \right)^2} + i \frac{\kappa}{4g} ~.
\end{equation}
We note that the definition of coherent BS angle $\theta$ is the same as in the lossless case, i.e. when $\kappa \rightarrow 0$.

Obtaining the system-bath transformation amplitudes, $\tilde{T}_{i k}$, from direct integration would be tedious.  Instead, we employ a general result which shows that any linear, dissipative two-mode transformation can be represented as a sub-system of a
four-mode unitary transformation \cite{Tischler:2018eh}.  
Furthermore, because the our evolution is passively linear and purely dissipative (i.e. no gain), the system-bath transformation amplitudes can be uniquely determined from the transformation matrix of the system modes alone.  

Consider the SVD of the system transformation matrix:
\begin{equation}
\tilde{\bm{T}} \equiv \left( \begin{array}{cc} \tilde{T}_{11} & \tilde{T}_{12} \\ \tilde{T}_{21} & \tilde{T}_{22} \end{array} \right) = \tilde{\bm{U}} \left( \begin{array}{cc} \lambda_1 & 0 \\ 0 & \lambda_2 \end{array} \right) \tilde{\bm{W}}~,
\end{equation}
where $\tilde{\bm{U}}$ and $\tilde{\bm{W}}$ are 2x2 matrices that represent lossless BS operations.  We necessarily have $\lambda_1,\lambda_2 \leq 1$ because the system is purely dissipative (i.e. no gain).  The evolution in Eq.~(\ref{eq:LBS_1}) can be written as \cite{Tischler:2018eh}
\begin{eqnarray}\label{eq:LBS_2}
\left( \begin{array}{c} \hat{a}_1(t) \\ \hat{a}_2 (t) \end{array} \right) &=&
 \tilde{\bm{U}} \left( \begin{array}{cc} \lambda_1 & 0 \\ 0 & \lambda_2 \end{array} \right) \tilde{\bm{W}} \left( \begin{array}{c} \hat{a}_1(0) \\ \hat{a}_2 (0) \end{array} \right) \nonumber \\
&&+ \tilde{\bm{U}} \left( \begin{array}{cc} \sqrt{1-\lambda_1^2} & 0 \\ 0 & \sqrt{1-\lambda_2^2} \end{array} \right) \left( \begin{array}{c} \hat{a}^\textrm{in}_3 \\  \hat{a}^\textrm{in}_4 \end{array} \right)~,~
\end{eqnarray}
where $\hat{a}^\textrm{in}_3$ and $\hat{a}^\textrm{in}_4$ are two orthogonal temporal bath modes;  the system-bath transformation amplitudes $\tilde{T}_{i3}$ and $\tilde{T}_{i4}$ can be obtained from the second line.  The physical meaning of this method is that if the temporal bath modes are defined appropriately, only two independent modes are needed to describe all the effects of the environment on the transduction.  We note that the temporal profile of $\hat{a}^\textrm{in}_3$ and $\hat{a}^\textrm{in}_4$ is unrelated to our analysis if we assume a vacuum bath.

To execute our scheme, two lossy BS are implemented by evolving the system according to Eq.~(\ref{eq:LBS}) twice, each with duration $\tau/2$.  In between the two BS, mode 1 is squeezed with strength $\gamma$.  Because there are two lossy BS, the environmental noise can be described by at most four temporal bath modes, i.e. $\ain_3$ and $\ain_4$ ($\ain_5$ and $\ain_6$) are responsible for the loss in the first (second) BS.  

Because $\tilde{\bm{T}}$ in Eq.~(\ref{eq:T_tilde}) is real, both the singular values ($\lambda_1$ and $\lambda_2$) and the matrices $\tilde{\bm{U}}$ and $\tilde{\bm{W}}$ are real.  If we decompose the mode operators into real and imaginary parts (quadratures), i.e. $\hat{a}_k=(\hat{q}_k+i\hat{p}_k)/\sqrt{2}$, they do not mix with each other.  In other words, each lossy BS is quadrature-diagonal for both system and bath modes.  Furthermore, the mode-1 squeezing at $t=\tau/2$ is also quadrature-diagonal.  As a result, the overall transformation (from $t=0$ to $\tau$) is quadrature-diagonal, i.e. 
\begin{eqnarray}\label{eq:LBS_x}
\xout_i &=& T^{xx}_{i1} \xin_1 + T^{xx}_{i2} \xin_2 + \sum_{k=3}^6 T^{xx}_{ik} \xin_k
\end{eqnarray}
where $i\in\{1,2\}$ and $x\in\{q,p\}$; we have extended the $\bm{T}$ matrix in Eq.~(\ref{eq:incomplete}) to include temporal bath modes.  

The transformation amplitudes are given by
\begin{eqnarray}
T^{qq}_{ik} &=& \gamma \tilde{T}_{i1}\tilde{T}_{1k} + \tilde{T}_{i2}\tilde{T}_{2k} \\
T^{pp}_{ik} &=& \frac{1}{\gamma} \tilde{T}_{i1}\tilde{T}_{1k} + \tilde{T}_{i2}\tilde{T}_{2k}
\end{eqnarray}
for $i=1,2$ and $k=1,2,3,4$, and
\begin{equation}
T^{qq}_{i5} = T^{pp}_{i5} = \tilde{T}_{i3}~,~T^{qq}_{i6} = T^{pp}_{i6} = \tilde{T}_{i4}~.
\end{equation}

For quantum memory write-in (one-way transduction from mode 1 to 2), our scheme starts by impedance-matching one quadrature by destructive interference.  We pick this to be the $q$-quadrature without loss of generality; this requires $T^{qq}_{22}=0$.  This can be achieved by choosing the mode-1 squeezing strength as $\gamma = -(\tilde{T}_{22})^2/\tilde{T}_{21}\tilde{T}_{12}$.  

To complete the transduction, the $p$-quadrature reflection noise should also be suppressed.  
We have discussed a strategy to first measure the $p$-quadrature of mode-1's output state, 
and then perform a conditional displacement on mode 2.  When there is no environmental noise, we have discussed that the conditional displacement strength $\eta_D$ should be the same as the unwanted QND strength, i.e. $\eta_D=\eta$.

On the other hand, in the lossy case the measurement outcome $p_1$ contains both the reflected quadrature noise and environmental noise.  Here the choice of $\eta_D$ is more subtle, because a full suppression of reflection noise might conversely enhance the environmental noise.  Therefore an optimal choice of $\eta_D$ should minimise the sum of these two noises.  We note that the conditional displacement is still assumed to be linearly proportional to the measurement outcome for simplicity; we leave it as an open question whether it is advantageous to use a more complicated non-linear dependence of the feedforward displacement on measurement outcome.

To obtain the optimal $\eta_D$, we first consider the total output state $\rho^\textrm{out}_\textrm{tot}$ of all system and bath modes after the overall transformation (\ref{eq:LBS_x}).  
After homodyne detection and conditional displacement, the output state at mode 2 is
\begin{eqnarray} 
\rho^\textrm{out}_2 &=& \int dp_1 \textrm{Tr}_{\backslash 2} \left\{ e^{-i\eta_D p_1 \hat{q}_2} \bra{p_1} \rho^\textrm{out}_\textrm{tot} \ket{p_1} e^{i\eta_D p_1 \hat{q}_2 }\right\} ~~\nonumber \\
&=&  \textrm{Tr}_{\backslash 2} \left\{ e^{-i\eta_D \hat{p}_1 \hat{q}_2} \rho^\textrm{out}_\textrm{tot} e^{i\eta_D \hat{p}_1 \hat{q}_2 }\right\}~. \label{eq:homo2}
\end{eqnarray}
In the second step, we have employed the identity
\begin{eqnarray}
\ket{p_1}\bra{p_1}\otimes e^{\pm i\eta_D p_1 \hat{q}_2} &=& e^{\pm i\eta_D \hat{p}_1 \hat{q}_2} \left(\ket{p_1}\bra{p_1}\otimes \hat{\mathbb{I}}_2 \right) \nonumber \\
 &=&  \left(\ket{p_1}\bra{p_1}\otimes \hat{\mathbb{I}}_2 \right)e^{\pm i\eta_D \hat{p}_1 \hat{q}_2}~,~~
\end{eqnarray}
where $ \hat{\mathbb{I}}_2$ is the identity of mode 2.  
The physical intuition is that applying a conditional displacement after a homodyne detection is equivalent to applying a QND gate before the homodyne detection.  

As such, the result of homodyne detection and post-processing can be accounted for by simply considering an \textit{effective} transformation, which an extra QND gate is applied on the output modes.  
The effective quadrature transformation remains in the form of Eq.~(\ref{eq:LBS_x}), i.e.
\begin{eqnarray}\label{eq:LBS_x2}
\xout_2 &=& \mathscr{T}^{xx}_{21} \xin_1 + \mathscr{T}^{xx}_{22} \xin_2 + \sum_{k=3}^6 \mathscr{T}^{xx}_{2k} \xin_k~,
\end{eqnarray}
but the transformation amplitudes are modified as
\begin{equation}\label{eq:LBS_sub}
\mathscr{T}^{qq}_{2k} = T^{qq}_{2k}~~,~~\mathscr{T}^{pp}_{2k} = T^{pp}_{2k} - \eta_D T^{pp}_{1k}~.
\end{equation}
We note that the output mode 1 can be neglected in this effective transformation (c.f. (\ref{eq:homo2})).

For the standard approach (where one just evolves once under the BS Hamiltonian for a fixed time), the effective transformation is also given by Eq.~(\ref{eq:LBS_x2}), except the parameters are replaced by $\gamma=1$ (i.e. no amplification at $t=\tau/2$) and $\eta_D=0$ (i.e. no measurement and post-processing).

We quantify the performance of the transducer by the total added noise (in units of quanta) in both quadratures,
\begin{equation}\label{eq:added_noise2}
\overline{\mathcal{N}}\equiv \mathcal{N}_q+\mathcal{N}_p
\end{equation}
where the added noise in $x$-quadrature is \cite{1982PhRvD..26.1817C}
\begin{equation}
\mathcal{N}_x \equiv \frac{1}{2}\sum_{k\geq 2}^6 \frac{(\mathscr{T}^{xx}_{2k})^2 \left\langle (\xin_k)^2 \right\rangle}{(\mathscr{T}^{xx}_{21})^2}~,
\end{equation}
for $x\in\{q,p \}$.
With this metric, the optimal displacement strength $\eta_D^\textrm{opt}$  should be that minimises $\overline{\mathcal{N}}$, i.e. 
\begin{equation}\label{eq:opt_eta}
\frac{\partial \mathcal{N}_p}{\partial \eta_D}\Big|_{\eta_D \rightarrow \eta_D^\textrm{opt}} =0~.
\end{equation}
We note that $\mathcal{N}_q$ does not depend on $\eta_D$.  

The total added noise in Eq.~(\ref{eq:added_noise2}) is not invariant under local transformation of the input state.  Specifically, because of the local squeezing and the effective QND gate, the added noise is generally different for the $q$- and $p$-quadratures.  If the input state is amplified in the more noisy quadrature, the added noise could be reduced \cite{1982PhRvD..26.1817C}.  Generally, for an initial phase sensitive amplification step that transforms the input mode as $\qin_1 \rightarrow \gamma_0 \pin_1$ and $\pin_1 \rightarrow \frac{1}{\gamma_0} \pin_1$, the added noise is modified as
\begin{equation}\label{eq:added_noise3}
\overline{\mathcal{N}}\rightarrow \frac{1}{\gamma_0^2}\mathcal{N}_q+\gamma_0^2\mathcal{N}_p ~.
\end{equation}
It is easy to find that $\overline{\mathcal{N}}$ is the minimum when the initial amplification is chosen as $\gamma_0^2 = \sqrt{\mathcal{N}_q/\mathcal{N}_p}$, then
\begin{equation}
\overline{\mathcal{N}}_{\min} = 2 \sqrt{\mathcal{N}_q \mathcal{N}_p}~.
\end{equation}

\add{We note that the definition of added noise in Eq. (89), which follows that in \cite{1982PhRvD..26.1817C} for quantifying noise of linear amplifiers, is given by taking the total fluctuations at the output that did not originate at the input, and dividing by the transmission probability.  In amplifier terminology, this corresponds to referring the output fluctuations back to the input.  We stress that reflected noise at the input contributes to this added noise.  
In the case of Fig.~\ref{fig:noise6}, because the BS interaction is applied for only $\tau_0/10$, the transmission amplitude is small.  Therefore, although the reflected input field is in vacuum, its contribution is still large on the scale of the weakly transmitted signal.  This translates to an added noise of $\approx 20$ quanta.}

Apart from added noise, the performance of a one-way transducer can also be quantified by its effective quantum channel capacity \cite{Weedbrook:2012fe}.  This metric is particularly useful when the transducer is applied in quantum communication, e.g. quantum memory transfer inside a quantum repeater.  For simplicity, we compute the quantum capacity, $\mathcal{Q}$, with a single use of channel and transmitting pure Gaussian state \cite{2001PhRvA..63c2312H, 2009PhRvL.102e0503P}; this quantity is a lower bound of the general quantum channel capacity.  $\mathcal{Q}$ can be calculated from the channel transmissivity \cite{Holevo:2007gn}, $\tau_\textrm{C}$, and the noise number $n_\textrm{C}$,
\begin{equation}\label{eq:Q}
\mathcal{Q} = \max \left\{0, \log \Big|\frac{\tau_\textrm{C}}{1-\tau_\textrm{C}} \Big| - G(n_\textrm{C}) \right\}~,
\end{equation}
where
\begin{equation}
G(n_\textrm{C}) \equiv (n_\textrm{C}+1) \log (n_\textrm{C}+1) - n_\textrm{C} \log (n_\textrm{C})~.
\end{equation}
For a one-way transduction from mode 1 to 2 that transforms the quadratures as in Eq.~(\ref{eq:LBS_x}) (effectively as in Eq.~(\ref{eq:LBS_sub}) because of measurement and feedforward), the effective channel transmissivity and noise number are given by
\begin{eqnarray}
\tau_\textrm{C} &=& \mathscr{T}^{qq}_{21}\mathscr{T}^{pp}_{21} ~,\\
n_\textrm{C} &=& \frac{\sqrt{\left(\sum_{k=2}^6 (\mathscr{T}_{2k}^{qq})^2 \langle (\qin_k)^2\rangle\right)\left(\sum_{k=2}^6 (\mathscr{T}_{2k}^{pp})^2 \langle (\pin_k)^2\rangle \right)}}{|1-\tau_\textrm{C}|}  \nonumber \\
&& -\frac{1}{2}~.
\end{eqnarray}
We note that when $\tau_C\leq 1/2$, the channel is anti-degradable \cite{Weedbrook:2012fe}, and Eq.~(\ref{eq:Q}) gives the exact channel capacity, i.e. $\mathcal{Q}=0$.

We note that although the above analysis focuses on the write-in process (transduction from mode 1 to 2), the readout process (transduction from mode 2 to 1) can be studied similarly after making two changes.  First, instead of destructively interfering the $q_2$ reflection (i.e. $T^{qq}_{22}=0$), the local squeezing between two BS should be adjusted to destructively interfere the $p_1$ reflection, i.e. $T^{pp}_{11}=0$.  Both conditions are simultaneously satisfied when there is no loss, which is a property of sQND.  However, this is not generally true in the lossy case.  Second, instead of using homodyne detection to remove unwanted quadrature noise, infinitely squeezed vacuum is injected into input mode 1. As usual, this is because we have assumed that mode 2 is less controllable than mode 1.

Without measurement and post-processing, it is not necessary to construct an effective transformation, and so the transduction is fully characterised by Eq.~(\ref{eq:LBS_x}).  The $x$-quadrature added noise in the readout scheme is given by
\begin{equation}
\mathcal{N}_x \equiv \frac{1}{2} \sum_{k\geq 3}^6 \frac{(T^{xx}_{1k})^2 \langle (\xin_k)^2\rangle}{(T^{xx}_{12})^2}~.
\end{equation}
We note that $(T^{qq}_{11})^2 \langle (\qin_1)^2\rangle =0$ due to the injected infinite squeezing.  The channel capacity can be computed by the effective channel parameters:
\begin{eqnarray}
\tau_\textrm{C} &=& T^{qq}_{12}T^{pp}_{12} ~,\\
n_\textrm{C} &=& \frac{\sqrt{\left(\sum_{k=3}^6 (T_{1k}^{qq})^2 \langle (\qin_k)^2\rangle\right)\left(\sum_{k=3}^6 (T_{1k}^{pp})^2 \langle (\pin_k)^2\rangle \right)}}{|1-\tau_\textrm{C}|}  \nonumber \\
&& -\frac{1}{2}~.
\end{eqnarray}
We have computed, but not shown, that the performance of readout is similar to that of write-in in Fig.~\ref{fig:noise6}.

\bibliographystyle{naturemag}
\pagestyle{plain}
\bibliography{interference_bib}

\end{document}